\documentclass[letterpaper, singlespace,twocolumn]{emulateapj}
\usepackage{natbib} 
\usepackage{graphicx}
\usepackage{bm}
\usepackage{hyperref}
\usepackage{subfigure}
\usepackage{amsmath}
\usepackage{amssymb}
\usepackage{wasysym}

\begin{document}

\title[Investigating the Consistency of Stellar Evolution Models with Globular Cluster Observations via the Red Giant Branch Bump]{Investigating the Consistency of Stellar Evolution Models \\with Globular Cluster Observations via the Red Giant Branch Bump}

\author{M. Joyce\altaffilmark{1} and B. Chaboyer\altaffilmark{1}}
\affil{Department of Physics and Astronomy, Dartmouth College, Hanover, NH 03755}

\begin{abstract}
Synthetic RGBB magnitudes are generated with the most recent theoretical stellar evolution models computed with the Dartmouth Stellar Evolution Program (DSEP) code. They are compared to the observational work of Nataf et al.,\ who present RGBB magnitudes for 72 globular clusters. A DSEP model using a chemical composition with enhanced $\alpha$ capture [$\alpha$/Fe] $ =+0.4$ and an age of 13 Gyr shows agreement with observations over metallicities ranging from [Fe/H] = $0$ to [Fe/H] $\approx-1.5$, with discrepancy emerging at lower metallicities.
\end{abstract}


\maketitle
\section{Introduction}
The evolutionary phase on the color--magnitude diagram known as the Red Giant Branch Bump (RGBB)  is a feature whose brightness is highly sensitive to the age and chemical composition of a stellar population. The bump is generated when the hydrogen-burning shell of an RGB star encounters the maximum depth reached by the convective envelope, which occurs early on the RGB. The chemical discontinuity causes a sudden availability of more fuel, resulting in a temporary drop in luminosity that is otherwise strictly increasing along the Red Giant Branch. Due to their sensitivity to the internal properties of stars, the RGBB luminosity and number count are excellent probes of the interior structure and mixing of stars.

The importance of the RGB bump for testing stellar evolution models has long been recognized.
Studies devoted to investigating the agreement between theory and observation of the RGBB date back to the work of \citet{Pecci}, who recognized that the bump provides an independent measure of cluster distance scale. Through luminosities of the bump and horizontal branch (HB) stars, 
{
\citet{Pecci} obtained values for the the magnitude difference between the bump and the horizontal branch, $\Delta M^{\text{HB}}_{\text{v}}$, which they used to calibrate the slope of the magnitude--metallicity relation.
}
A few years later, \citet{CS97} presented theoretical RGBB and zero-age horizontal branch (ZAHB) luminosities over a range of typical globular cluster metallicities, comparing theoretical values of $\Delta V^{\text{bump}}_{\text{HB}}$ with observations. \citet{Ferraro} presented a catalog of 61 globular clusters (GCs) and located and analyzed the RGBB in 47 of them.\ 
 Soon after, \citet{Bergbusch} discussed the failure of their isochrone population function (IPF) software to generate models consistent with RGBB observations, recording a discrepancy of ${\sim} 0.25$ mag. 
\citet{Riello} presented the magnitude difference between the luminosity function (LF) of the RGBB and the HB, as well as the star counts in the bump region for a sample of 54 GCs. They found qualitative agreement between theory and observation, but reported a small average discrepancy between the number of predicted versus observed star counts, varying with metallicity.

It is important to note that throughout the study of the RGBB, astronomers have employed many different techniques when comparing observations with theory. In particular, the observational quantity chosen for comparison with models varies across the literature. 
This quantity is typically expressed as a magnitude difference between the RGBB and another distinct observational feature, such as $\Delta V^{\text{HB}}_{\text{bump}}$ \citep{CS97}, the difference between the bump and the main sequence turn-off (MSTO) $\Delta V^{\text{MSTO}}_{\text{\text{bump}}}$
\citep{Cas}, or the difference between the bump and a given point along the main sequence (MS) $\Delta V^{\text{MS}}_{\text{bump}}$ \citep{Troisi}.  

The first portion of our analysis in this work involves adopting an estimated uncertainty in the GC distance modulus to obtain the RGBB luminosity, rather than using the difference between the RGBB and another feature. 
We note that other groups have proceeded differently, in many cases specifically to avoid the adoption of an uncertain GC distance scale. However, we think this method is a better test of the RGBB magnitude {\it because} it does not depend upon the properties of stellar models in other evolutionary phases. This is discussed further in section 6.

In their earlier work, \citet{Chab06} concluded that models generated with the Dartmouth Stellar Evolution Program (DSEP) code were consistent with the GC RGBB observations of \citet{Zoc}, whose observational sample included data on 28 GCs taken with {\it HST}. In 2010, Di Cecco et al.\ disagreed with this consistency and published a magnitude difference of  ${\sim} 0.4$, where $\Delta V_{\text{bump}} = V_{\text{HB}} - V_{\text{RGBB}}$ was observed to be larger than predicted by the { BaSTI models of \citet{Piet06}}. Their observational sample was taken with ground-based instruments and included 62 GCs, of which 40\% showed discrepancies of $2\sigma$ or more \citep{Di}. They noted that the discrepancy increased in metal-poor GCs.  
{ 
A year later, \citet{Cas} conducted the first study ever to adopt a $\Delta V_{\text{bump}}^{\text{MSTO}} $ definition using the magnitude difference between the bump and the MSTO. Their study also implemented accurate GC dating so as to remove the age-dependence of the RGBB brightness
in order to account for the age dependency of both the RGBB and the MSTO.
}
They concluded that the  BaSTI models \citep{Piet06} under-predicted observed bump magnitudes as well, but this time by an average of ${\sim} 0.2$ magnitudes. 

Since then, observations have continued to improve, and after the publication of Nataf et al.'s 72-cluster survey, the question of whether DSEP models remain consistent with observation naturally arises. In this paper, we present a comparison between the most recent empirical bump magnitudes \citep{Nataf} and the magnitudes predicted by stellar models generated with DSEP.   \looseness=1 

\section{Models}

For a detailed discussion of the DSEP code, we refer the reader to \citet{Chab06} and \citet{Dotter}, wherein the internal workings of the code have been thoroughly described. Major adjustments to the DSEP models since they were last compared to RGBB magnitudes in 2006 include updates in nuclear reaction rates \citep{Adel, Marta}, the use of more sophisticated equations of state \citep{FreeEOS}, and improved surface boundary conditions from the PHOENIX model atmospheres \citep{Phoenix}. 
 The most illuminating way to illustrate the effects of these improvements is to generate a new grid of models using the same composition parameters as were used in the grid from \citet{Chab06}, hereafter BC2006.

For our investigation, suites of masses with various compositions are evolved, and evolutionary tracks whose RGBB center points occur at 11 Gyr and 13 Gyr (within 1\%) are selected. 
The RGBB bolometric magnitude, effective temperature, precise age, and surface gravity at the center of the RGBB are extracted from tracks of designated mass with custom software. 

{
It is conventional to determine the RGBB magnitude from isochrones interpolated from tracks. However, for the bulk of our analysis, we elect to determine the RGBB magnitude from individual stellar tracks whose masses are tuned to give the desired RGBB age.
We would expect the magnitudes produced directly from stellar tracks to be more accurate than interpolated magnitudes because the isochrone code only generates a few hundred points along the red giant branch, whereas individual stellar tracks typically contain a few thousand points in this region. Consequently, the isochrone code does not resolve the RGBB region as well as individual stellar tracks.
To improve our understanding of the numerical uncertainty, which is distinct from the physical uncertainty, we compare the RGBB magnitudes derived by each method.
}

\begin{figure} 
\centering
\includegraphics[width=\linewidth]{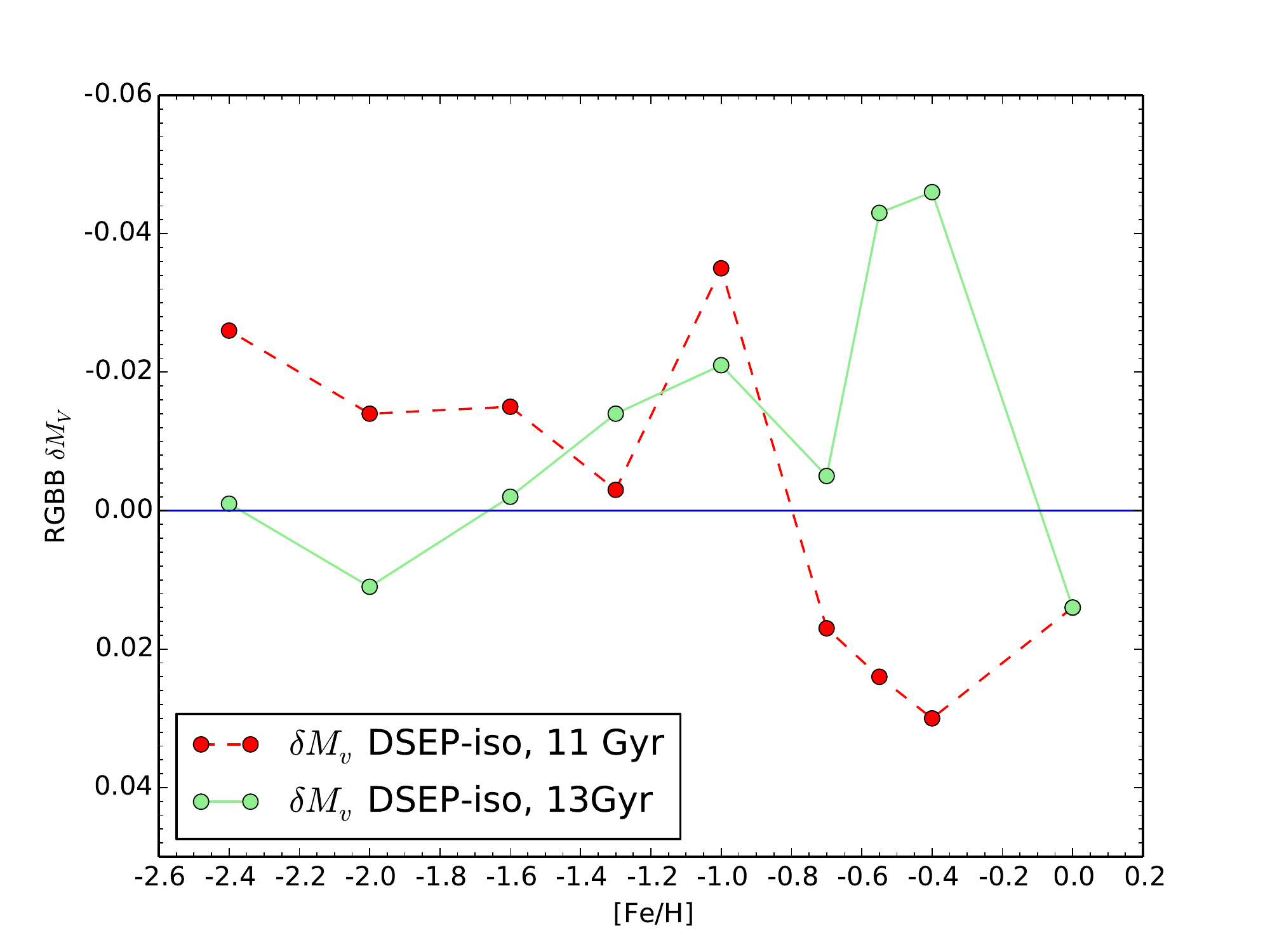}
\caption{The differences in magnitude for the RGBB as calculated directly from stellar evolution tracks (DSEP) and tracks interpolated using an isochrone code,
$M_{V,\text{DSEP}} - M_{V,\text{isochrone}}$, are shown as a function of metallicity. Calculations for both the 11 Gyr and 13 Gyr model grids are presented.}
\label{DV}
\end{figure}

Figure \ref{DV} shows the difference in RGBB magnitudes generated with the isochrone code versus those derived from DSEP's single-star evolutionary track 
 for stars tuned to ages ${\sim}11$ and ${\sim}13$ Gyr. 
The isochrone code interpolates among a number of stellar evolutionary calculations for stars with different masses but identical compositions.
The difference is indicative of the interpolation error in our isochrone generation code and/or the non-identical ages of the stellar tracks. We find an RMS difference between the two methods of ${\sim}0.02$ magnitudes. This value is indicative of the numerical uncertainty in our RGBB magnitudes.

To determine the RGBB magnitude from stellar tracks, the luminosity function for a given stellar model is calculated by assuming a very large number of stars is being born at a constant rate over the age range of interest on the RGB.  
{
A small issue with this method arises from the fact that models evolve a single-mass star over a span of ages, whereas real globular clusters contain stars of the same age with slightly different masses. 
For example, the ages that bound the RGBB region of a model with an age of 13 Gyr, [$\alpha$/Fe]=$+0.4$ and [Fe/H]=$-0.7$ 
are 12.98 Gyr and 13.05 Gyr. This relatively small age difference does not significantly impact the determination of the RGBB luminosity.
}

The conversion from theoretical luminosities and temperatures to observed magnitudes and colors is done using the \citet{VC} color tables. The V-band magnitudes after conversion are presented as the final synthetic RGBB magnitudes.
 Models are computed over metallicities ranging from $[{\rm Fe}/{\rm H}]=0$ to $-2.4$ for two values of $\alpha$-enhancement and RGBB ages of 11 Gyr and 13 Gyr for each composition.

\begin{figure} 
\centering
\includegraphics[width=\linewidth]{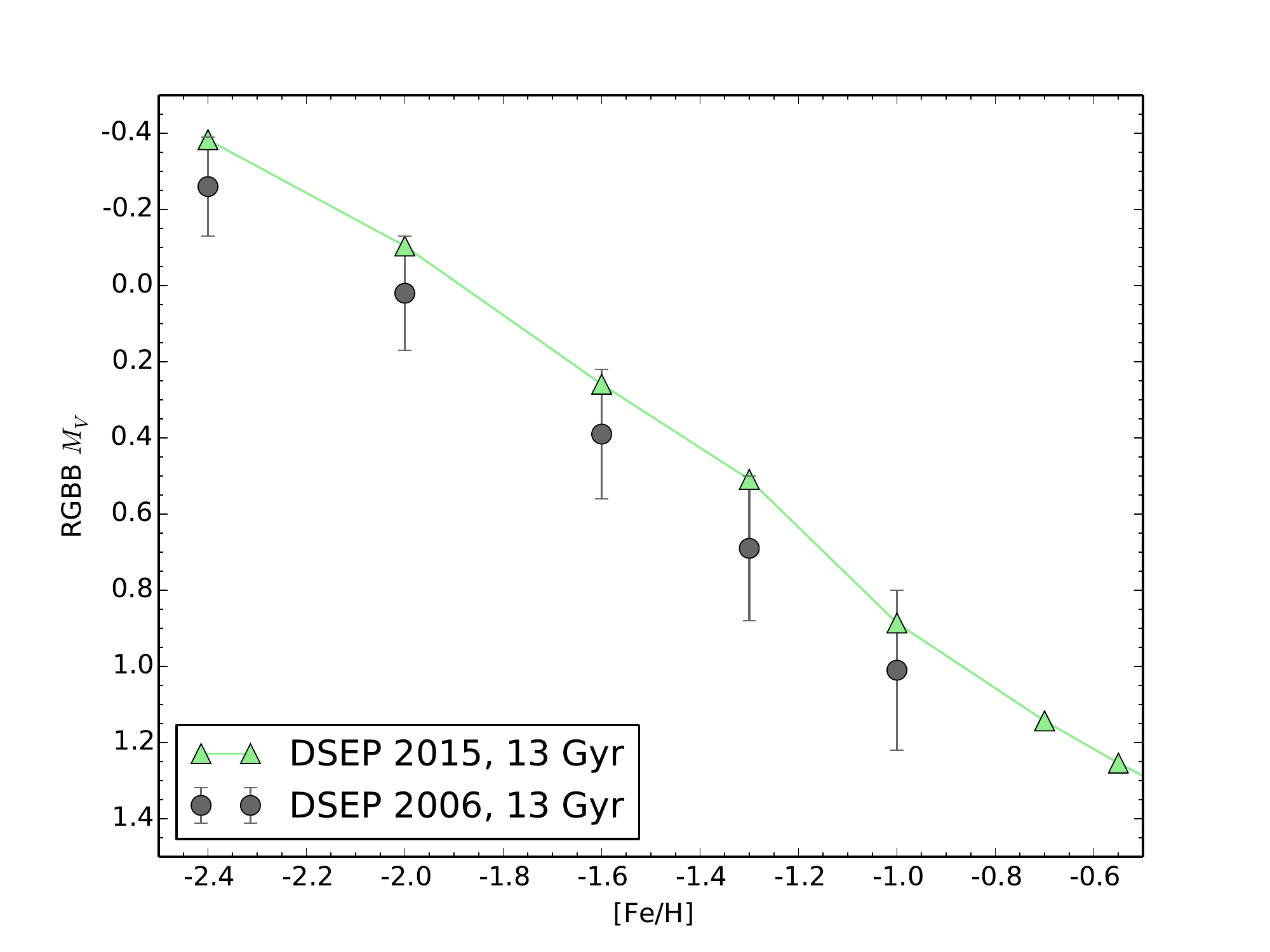}
\caption{RGBB magnitudes predicted by the BC2006 DSEP models are compared to the 2015 models. The models have [$\alpha$/Fe] = $+0.4$.}
\label{comp}
\end{figure}

The 2015 bump magnitudes are presented against the BC2006 set (which were only computed for [Fe/H] $\le-1.0$) as a function of [Fe/H] in Figure \ref{comp}, where the 68\% confidence limits (one standard error) of BC2006 are used as uncertainties for the 2006 models. 
The 2006 and 2015 DSEP models are in good agreement across the metallicity spectrum, with the 2015 magnitudes falling within the uncertainties reported by BC2006 for all points. In general, the 2015 models produce slightly brighter magnitudes. A plausible candidate for this difference is that the best known value of the $^{14}$N$(\text{p},\gamma) ^{15}$O nuclear reaction rate \citep{Marta} has changed by a factor of two, while BC2006 assumed only a 15\% uncertainty in this rate. 
{
Evidence in favor of this suggestion is provided by \citet{Piet10}'s discovery that a shift in magnitude of similar order and in the same direction resulted from the adjusted $^{14}$N$(\text{p},\gamma) ^{15}$O reaction rate.}
We assume that the models using improved physics provide more accurate values and proceed using the most updated version of DSEP.

\subsection{Comparison to Other Stellar Models}

\begin{figure} 
\centering
\includegraphics[width=\linewidth]{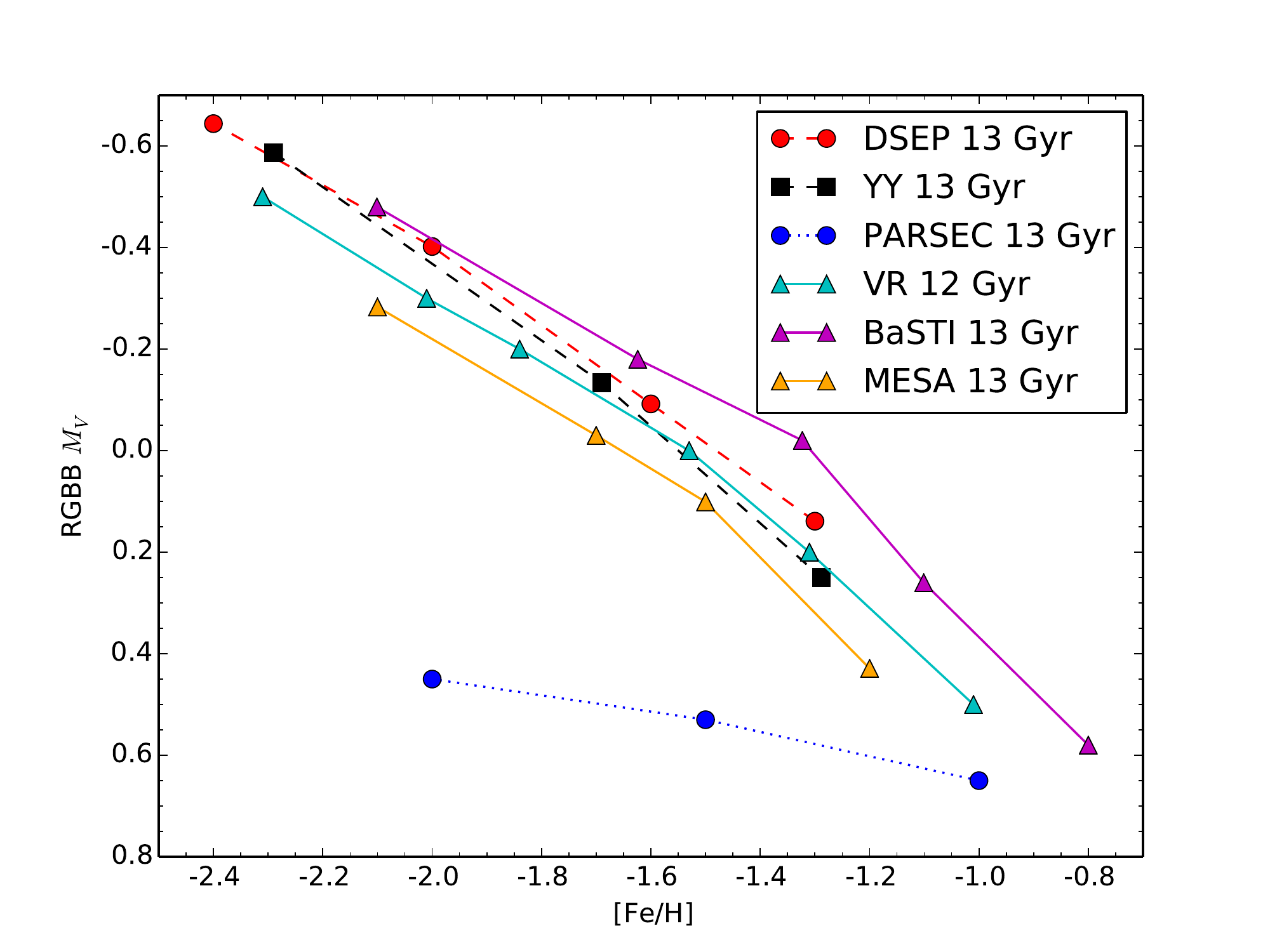}
\caption{RGBB magnitude as a function of [Fe/H] for different stellar models at the scaled solar composition ([$\alpha$/Fe] = 0.0).}
\label{othermodels00}
\end{figure}

\begin{figure} 
\centering
\includegraphics[width=\linewidth]{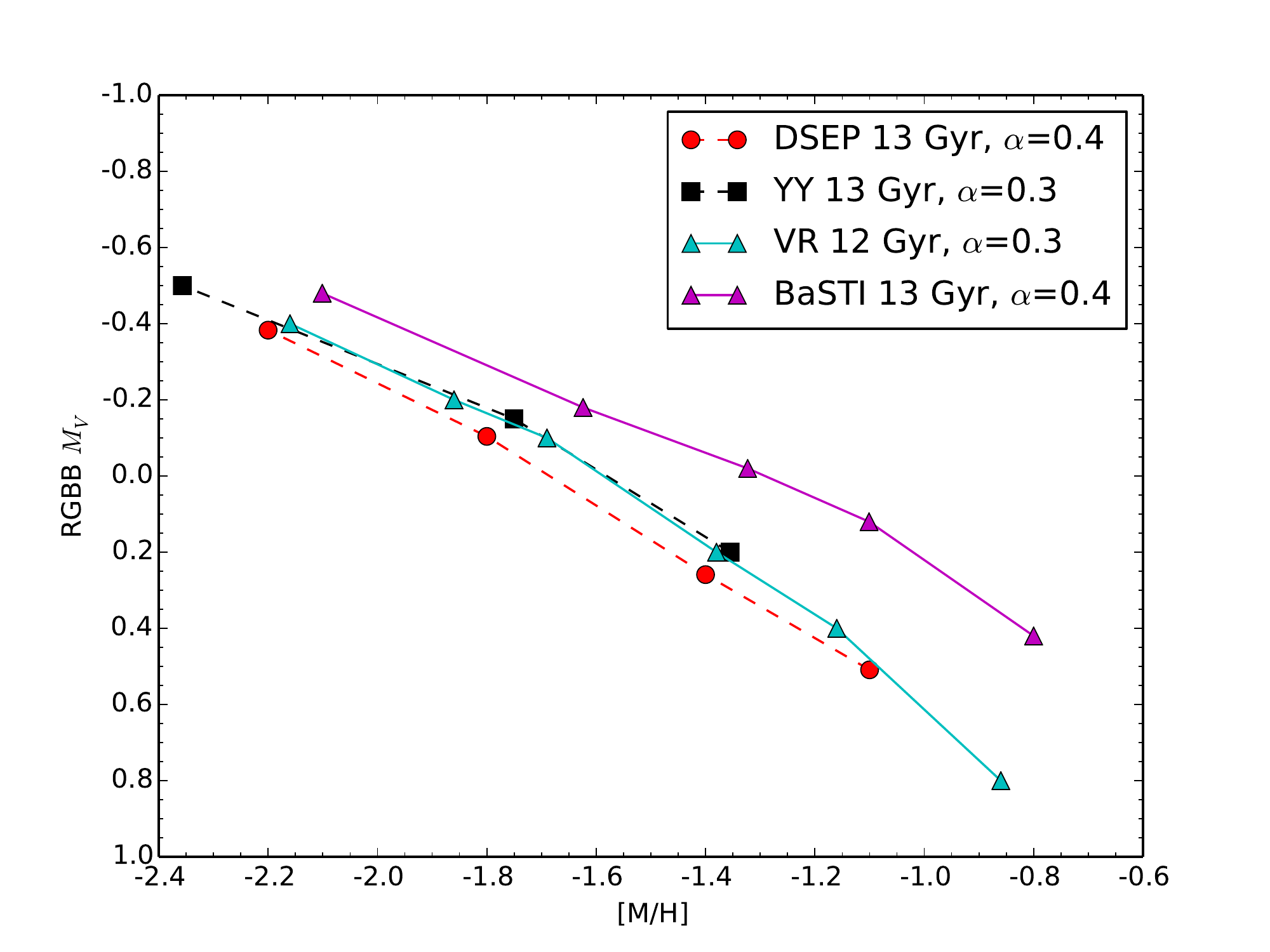}
\caption{RGBB magnitude as a function of global metallicity [M/H] for different $\alpha$-enhanced stellar models.}
\label{othermodels04}
\end{figure}

We next examine the stellar evolution models of other groups whose luminosity functions are easily available online. We compare with the Yonsei--Yale (YY) models of \citet{YY}, the Victoria--Regina (VR) models of \citet{VDB2006}, the PARSEC models of \citet{CMD}, and the BaSTI models of \citet{BaSTI}. In addition, we generate independent stellar models using the publicly available MESA stellar evolution code of \citet{MESA}. The MESA stellar tracks are tuned to an RGBB age of 13 Gyr in the same manner as the DSEP models.
The comparison between the DSEP magnitudes and those of the other models is shown in figures~\ref{othermodels00} and \ref{othermodels04}. The synthetic data
{ used in the investigations of \citet{Cas} and \citet{Di} are generated with the BaSTI luminosity functions. }

In Figure~\ref{othermodels00}, models with scaled solar compositions are compared. 
The magnitudes generated with DSEP agree best with the models of YY (this is to be expected for historic reasons) and VR, and with the BaSTI models at the lowest metallicities.

In Figure \ref{othermodels04}, the comparison is done for compositions of similar $\alpha$-enhancement and ages: the DSEP curve has $\alpha=+0.4$, the YY curve has $\alpha=+0.3$, the VR curve has $\alpha=+0.3$, the BaTSI curve has $\alpha=+0.4$, 
 and in the case of the PARSEC and MESA curves, { only models with scaled solar composition are available.}
The models are presented in terms of global metallicity to account for differences in $\alpha$-enhancement. The $\alpha$-enhanced models show similar trends to those with scaled solar compositions. In all cases, the BaSTI models predict the brightest RGBB.

 All of the models, with the exception of PARSEC's, agree within a span of roughly 0.2 magnitudes. Errors of this order are potentially attributable to differences in the microphysics implemented by each group. According to \citet{YY}, molecular diffusion is not included in the YY model calculations, and a gray model atmosphere is used for boundary conditions. The same is true for the VR models \citep{VDB2006}. In the case of PARSEC \citep{CMD}, a gray model atmosphere is used, but microscopic diffusion, which includes gravitational settling, is included. The DSEP models, on the other hand, use PHOENIX model atmosphere boundary conditions and include diffusion of helium and heavy elements.
{
We note the obvious disagreement between the PARSEC models and those of the other groups. The cause for discrepancy is not immediately clear.}

We conduct a brief experiment by recomputing a DSEP RGBB test magnitude with diffusion turned off in our code.
{
In general, the mixing length of the star is calibrated by a solar model; however, when the input physics of the model are changed, the mixing length must be recalibrated. We recompute the diffusion-free RGBB magnitude at this rescaled mixing length, and find that this accounts for a change of $\sim0.1$ magnitudes from the value computed with diffusion turned on.
}
 This is of the order necessary to push the DSEP curve just below the VR and YY curves, or into better alignment with the BaSTI curve. 
In addition, we examine the impact of using the rescaled mixing length independently of diffusion, as well as testing the effects of using a gray atmosphere and using less recent nuclear reaction rates. These changes are not found to have significant impacts; they each result in magnitude shifts of $\le 0.03$ individually. The insignificance of these changes has been documented in the more detailed work of previous authors \citep{CDS,MRR}. It is possible that a combination of microphysical differences---especially in diffusion---could account for differences within the $\approx0.2$ magnitude range described above.

It is not immediately obvious what causes the small differences among the various stellar evolution codes, but the synthetic magnitudes generated with DSEP reflect the literature consensus among other models. The true test of a model's robustness, however, is how well it fits to real data. We hence defer to an examination of DSEP's fit to observations in later sections.

\break
\section{Observational Data}
Before comparing the synthetic data to observations, it is instructive to intercompare observational data sets from the two large RGBB studies to which DSEP models will be compared. \citet{Zoc} (Z1999) used HST to obtain RGBB magnitudes of 28 globular clusters. More recently, \citet{Nataf} (N2013) obtained the RGBB magnitude of 72 clusters with HST.
N2013 determined $ V_{\text{bump}}$ by log-integrating the luminosity functions of cluster red giant stars on either side of the RGBB and measuring the point at which the two linear fits separate (a standard method). The Z1999 observations used HST's Wide Field and Planetary Camera 2 (WFPC2) CCD. The N2013 observations used ACS, the current CCD, for 55 GCs in their set and WFPC2 for 17 GCs. 
 
\begin{figure}
\centering
\includegraphics[width=\linewidth]{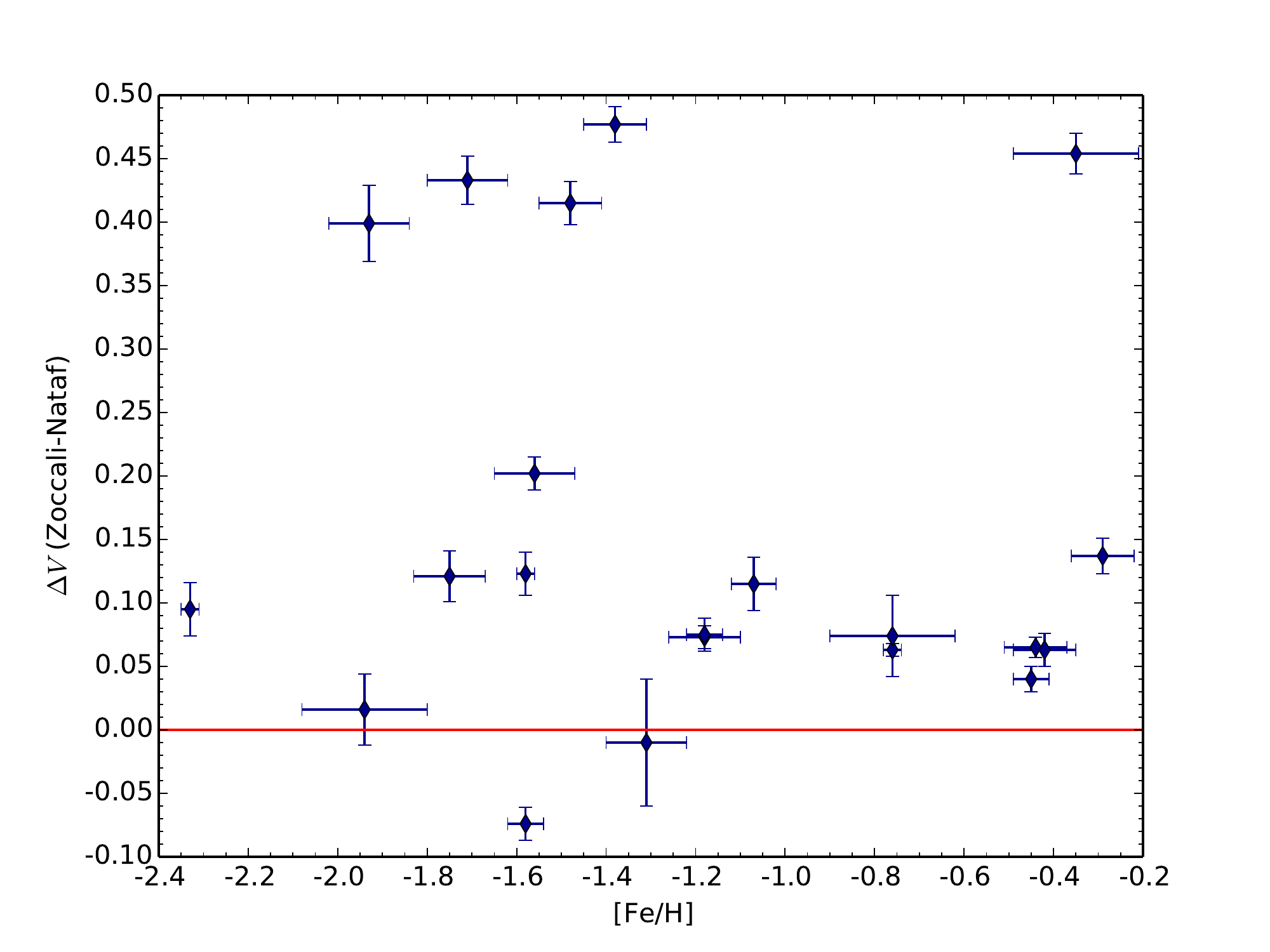}
\caption{Bump magnitude observations made by Z1999 and N2013 are compared. The magnitudes of N2013 are subtracted from the magnitudes of Z1999 ($V_{Z1999} - V_{N2013}$) and shown as a function of the N2013's [Fe/H] values. }
\label{glob}
\end{figure}

We compare the intersecting GC observations from Zoccali et al.\, (Z1999) and Nataf et al.\ (N2013) in Table \ref{table1}. 
Figure \ref{glob} gives the differences in bump magnitudes reported by both groups. It is found that the magnitudes reported by Z1999 are systematically fainter than those reported by N2013.

It should be noted that N2013 subdivide their data into ``silver" and ``gold" samples, where the gold data are regarded with higher confidence. The gold sample contains data from both ACS and WFPC2. They acknowledge that Z1999 have already examined their WFPC2 GC observations in detail, but do not make an explicit comparison between their data and Z1999.

We note that detection of the bump becomes more difficult at low metallicities because the number of stars decreases with metal depletion. The reason for this decrease is twofold: Firstly, the maximum depth of the convection zone does not reach as deeply in metal-poor stars, causing a smaller discontinuity in composition and thus a smaller luminosity change. The result is a less prominent bump in metal-poor stars than we would see in metal-rich ones. 
 Secondly, metal-poor stars encounter the discontinuity in composition at brighter luminosities, where the evolutionary timescales are shorter. This means there will be fewer stars in the bump region of metal-poor GCs.
Hence, we may anticipate that the observations would vary more at lower [Fe/H] values.

These considerations given, there are still five GCs that display an observational discrepancy of more than $m_V=\nobreak0.2$ magnitudes.
 This is especially noteworthy in the case of those clusters (marked with $\dag$ in Table \ref{table1}) for which the magnitude values in both papers were derived from the same WFPC2 data. Characteristic values of the error bars of the two data sets (which are added in quadrature to obtain the error bars in Figure \ref{glob}) are ${\sim} 0.03$ for Z1999 and ${\sim} 0.01$ for N2013.  A brief investigation of the observational discrepancy is conducted with a $\chi^2$ test using the error bars of both observing groups. The test reveals dramatic inconsistency, yielding a reduced $\chi^2$ of 34. It is unclear why there is such a striking difference,
 but the fact that there exist significant discrepancies among measurements taken from the same WFPC2 GCs may point to differences in the methodology for estimating the RGBB brightness.
This suggests that the observational error bars are underestimated, or some points are in error.

\begin{figure} 
\centering
\includegraphics[width=\linewidth]{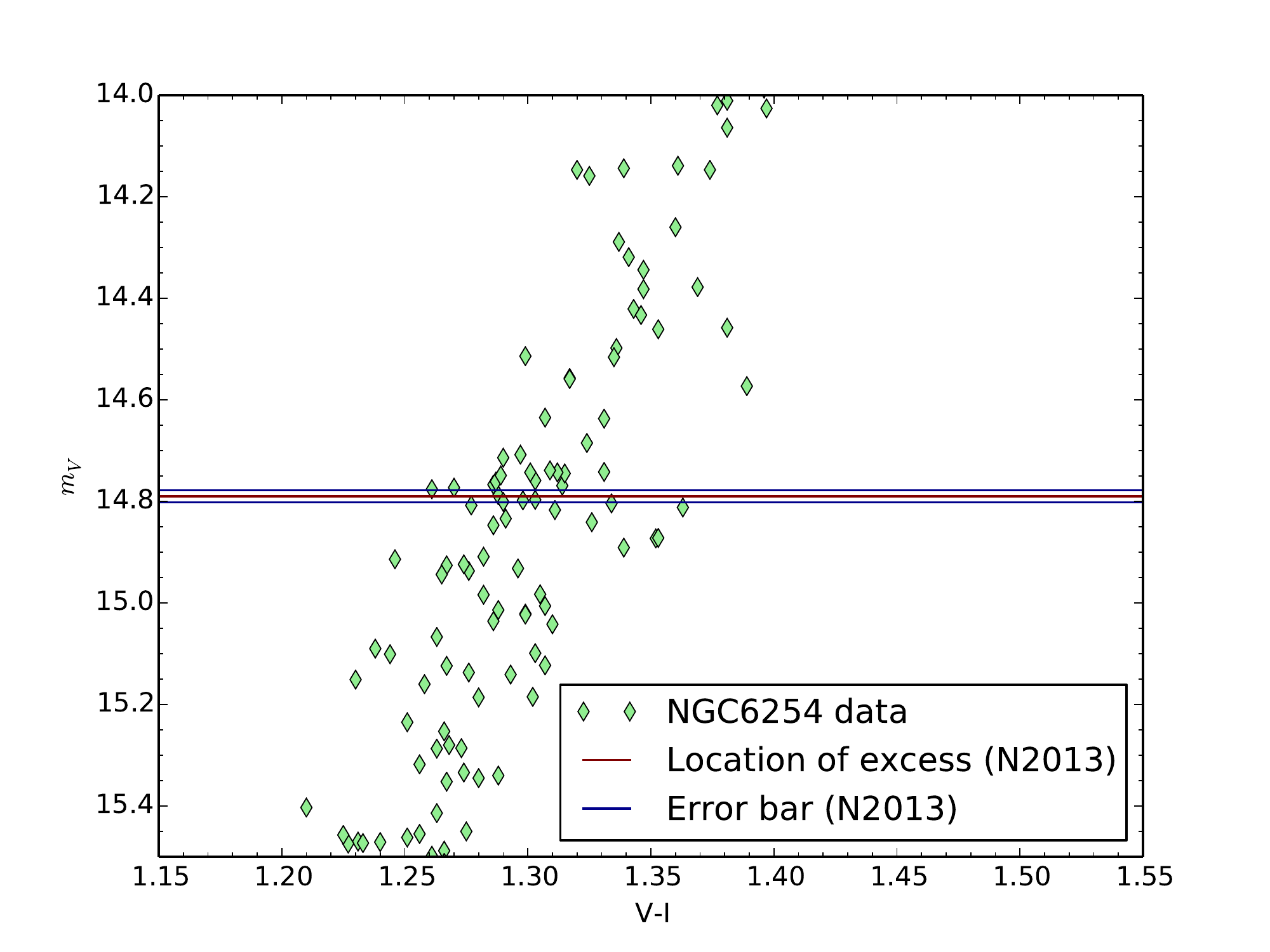}
\includegraphics[width=\linewidth]{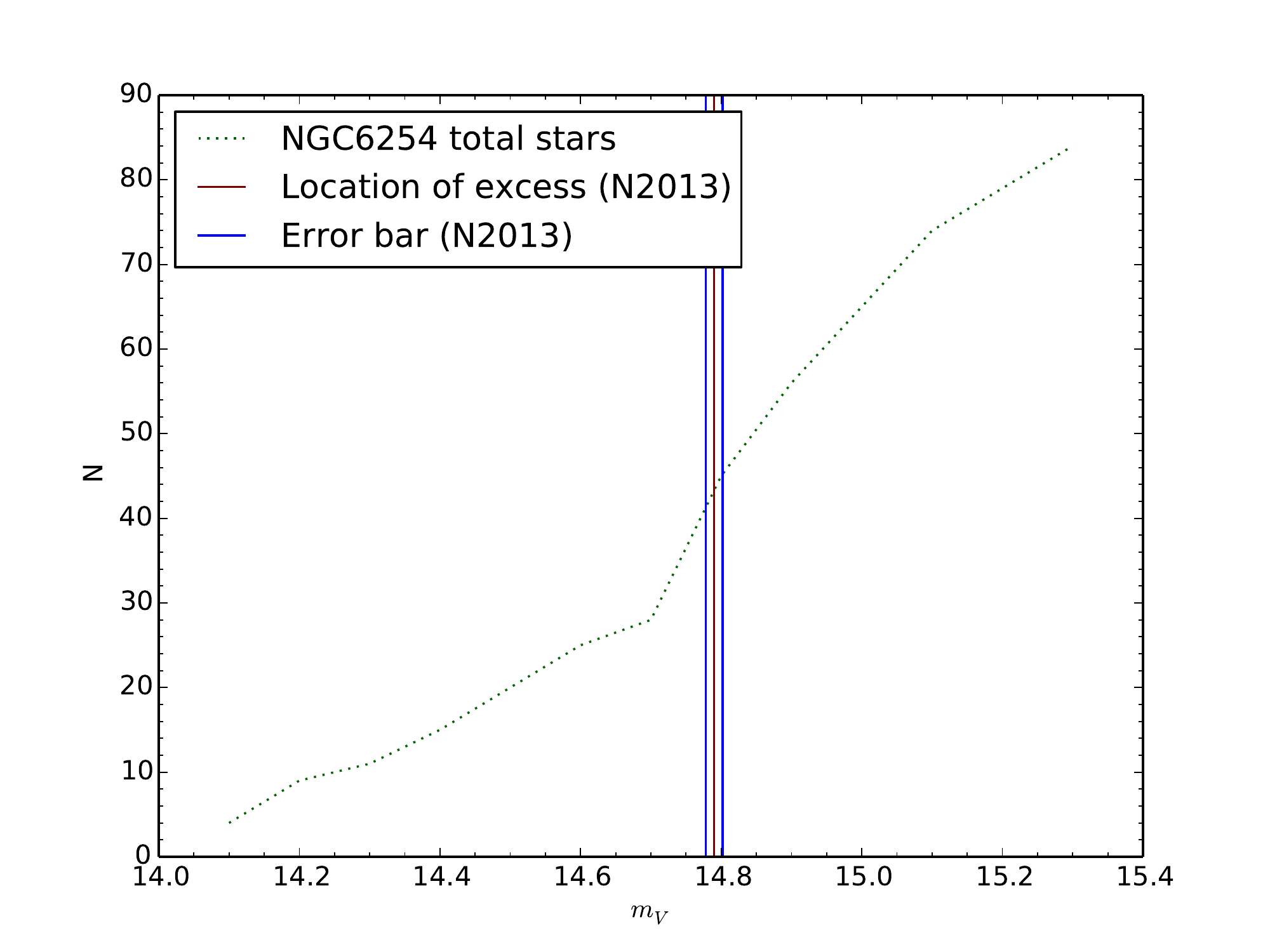}
\caption{TOP: A color--magnitude diagram constructed from observational data on NGC6254, with N2013's reported luminosity and error bars. BOTTOM: A cumulative luminosity function is shown with indicators as above.}
\label{Q}
\end{figure}

Probing further, we examine the data for one of N2013's clusters directly. We choose NGC6254, as this is the cluster that is most discrepant with our models (see sections 4 and 5). 
Figure \ref{Q} shows the raw data from the ACS globular cluster treasury program \citep{Sara} in two forms with N2013's reported magnitude and error bars superimposed. The top panel shows a color--magnitude diagram and the bottom panel shows a cumulative luminosity function.

 For NGC6254, N2013 quotes $V_\text{bump} = 14.79 \pm 0.012$. Though it is clear that 
the number of stars drops off dramatically around V=14.7 in Figure \ref{Q}, it is difficult to see an excess of stars at precisely V=14.79, and we are skeptical of such a small error bar on $V_{\text{bump}}$.
 A brief survey of the literature also reveals lower NGC6254 bump magnitudes more consistent with our predictions. In particular, \citet{Pol} provide a bump magnitude of $V_{\text{RGBB}} = 14.57 \pm 0.10$. Since Pollard et al.\ had wide field ground-based photometry with approximately double the number of stars in the RGB region, their value should be more accurate than the one reported by N2013.

Similarly, we note that for the very metal-poor cluster NGC6341 (M92, with [Fe/H] = $-2.35$), the luminosity function of \citet{Paust} yields a bump magnitude of $14.5 \pm 0.1$ mag, while Nataf et al. find $V_{\text{RGBB}} = 14.67 \pm 0.013$. \citet{Paust} combined wide field ground-based data with HST images, and their RGB has approximately double the number of stars as used by \citet{Nataf}.

Together, the disagreement between Z1999 and N2013 for many clusters and the more detailed investigation of the RGBB magnitude for NGC6254 suggest that N2013 may have underestimated their errors.
With these considerations in mind, we proceed using N2013 as the empirical basis for comparison with DSEP models and observe a note of caution regarding their error bars. 

\begin{table} 
\centering 
\caption{RGBB Magnitudes from Z1999 and N2013.}
\begin{tabular}{ l  l l l }  
\hline\hline
 { GC} & { Z1999 } & { N2013 } & { [Fe/H]} \\ \hline
NGC104 & 14.57 & 14.507 & -0.76 \\ 
NGC1851 & 16.16 & 16.087 & -1.18	\\ 
NGC1904  & 16.0 & 15.877 & -1.58	\\ 
NGC2808 & 16.31 & 16.235 & -1.18	\\ 
NGC5634 $ \dag $ & 17.77 & 17.371 &-1.93	\\ 
NGC5824 $ \dag $ & 18.1 & 18.084 &-1.94	\\ 
NGC5927 & 17.37 & 17.233 &-0.29	\\ 
NGC6093  $ \dag$  & 16.12 & 15.999 &-1.75	\\ 
NGC6139 $\dag $ & 18.3 & 17.867 &-1.71	 \\ 
NGC6205 & 14.7 & 14.774 &-1.58	 \\ 
NGC6235  $ \dag $ & 17.24 & 16.763 &-1.38	 \\ 
NGC6284 $ \dag $& 17.36 & 17.37 &-1.31	 \\ 
NGC6356 $ \dag $ & 18.53 & 18.076 &	-0.35 \\ 
NGC6362 & 15.6 & 15.485 &-1.07 \\ 
NGC6388 $ \dag $ & 17.69 & 17.65 & -0.45  \\ 
NGC6441 & 18.46 & 18.395 & -0.44 \\ 
NGC6624 & 16.68 & 16.617 & -0.42 \\ 
NGC6652 & 16.44 & 16.366 & -0.76 \\ 
NGC6934 & 16.85 & 16.648 & -1.56 \\ 
NGC6981 & 17.13 & 16.715 & -1.48 \\ 
NGC7078 & 15.41 & 15.315 &-2.33 \\ \hline
\end{tabular}
\tablecomments{The apparent magnitudes $m_V$ of the RGBB as given by Z1999 and N2013 are compared. The metallicities presented are those determined by N2013. Data on clusters marked with $\dag$ were collected with WFPC2 in both the Z1999 and N2013 samples.\\}
\label{table1}
  \end{table}

\section{Comparison Between Theory and Observations}

\begin{figure}	
\centering
\includegraphics[width=\linewidth]{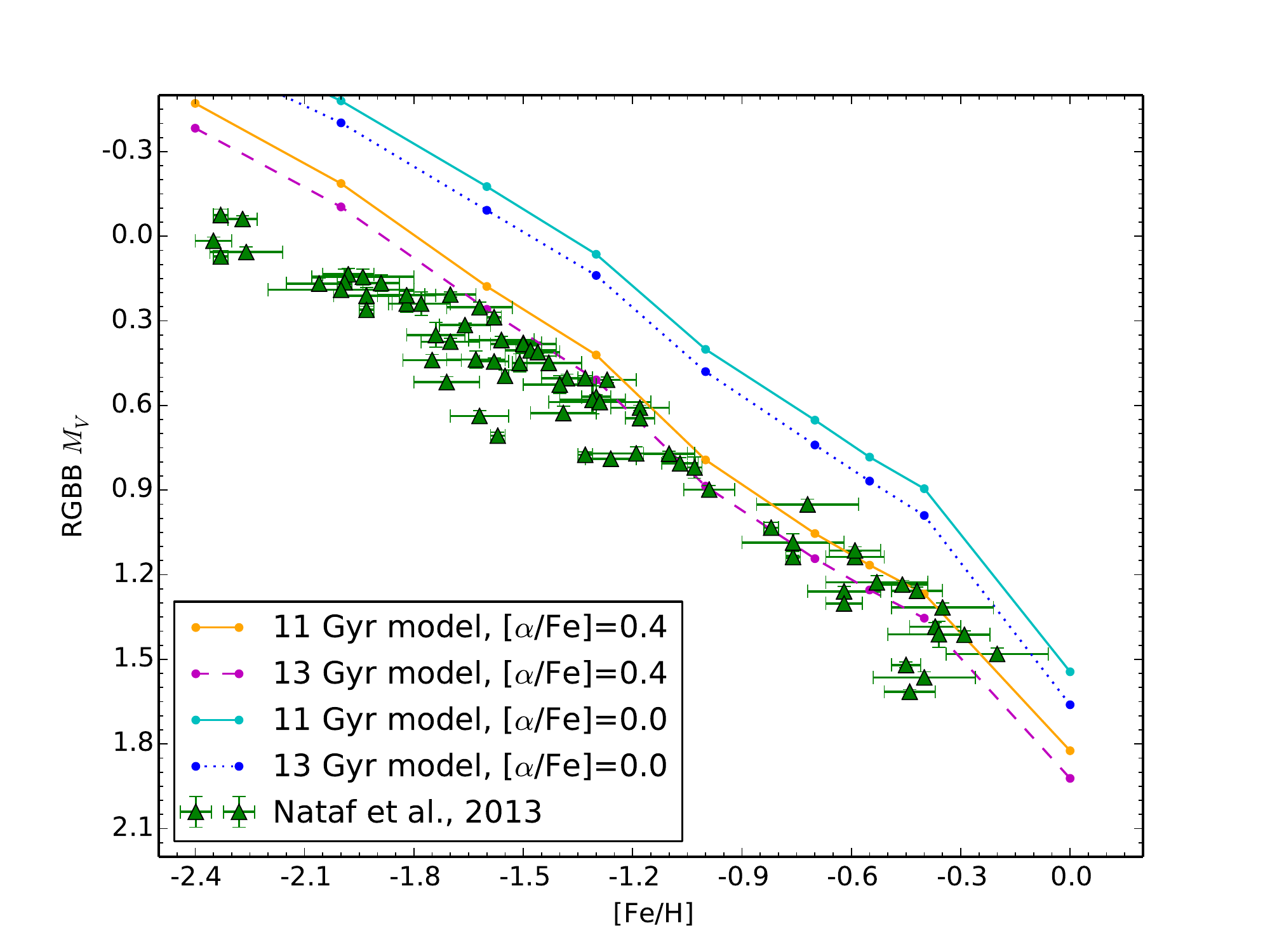}
\caption{Observations of N2013 are plotted against five sets of DSEP models over the metallicity range [Fe/H]=(-2.4,0). The V-magnitude of the RGBB is shown as a function of [Fe/H]. Uncertainties in the observed distance moduli are not pictured.}
\label{NJ}
\end{figure}

The DSEP model suite used in this analysis consists of magnitudes generated from evolutionary tracks tailored to 11 and 13 Gyr. The models are computed for both a scaled solar composition and an $\alpha$-enhancement of $+0.4$. 
N2013's observed RGBB magnitudes are compared to our models in Figure \ref{NJ}. For this comparison, we adopt the distance modulus to each cluster reported by N2013. All 72 GCs are shown. 

{
Both the 11 and 13 Gyr, $\alpha$-enhanced models tend to intersect the brighter boundary of the empirical distribution, 
for [Fe/H]$>-1.5$. For metallicities below [Fe/H] $=-1.5$ (hereafter referred to as the``cutoff point"), the data and models rapidly diverge.
 }
This trend is somewhat counterintuitive given that the BC2006 theoretical uncertainties are the smallest for low metallicities. 
A more detailed investigation into the statistics of the observational distribution suggests that some of the most discrepant GCs are outliers---mathematically speaking---and we probe the possible physical reasons for this later. 

Overall, a 13 Gyr, $\alpha$-enhanced model is found to provide the best ($\chi^2$-minimized) fit to the full sample. We also assess our fit to  N2013's higher-confidence ``gold" sample, which consists of 48 GCs, and the same set of DSEP models. The same 13 Gyr, $\alpha$-enhanced model is a better fit to the gold sample alone than it is to the entire sample; however, the fit still veers from the observations at low metallicities. We elaborate on the formal goodnesses of fit in the next section.

\begin{figure}	
\centering
\includegraphics[width=\linewidth]{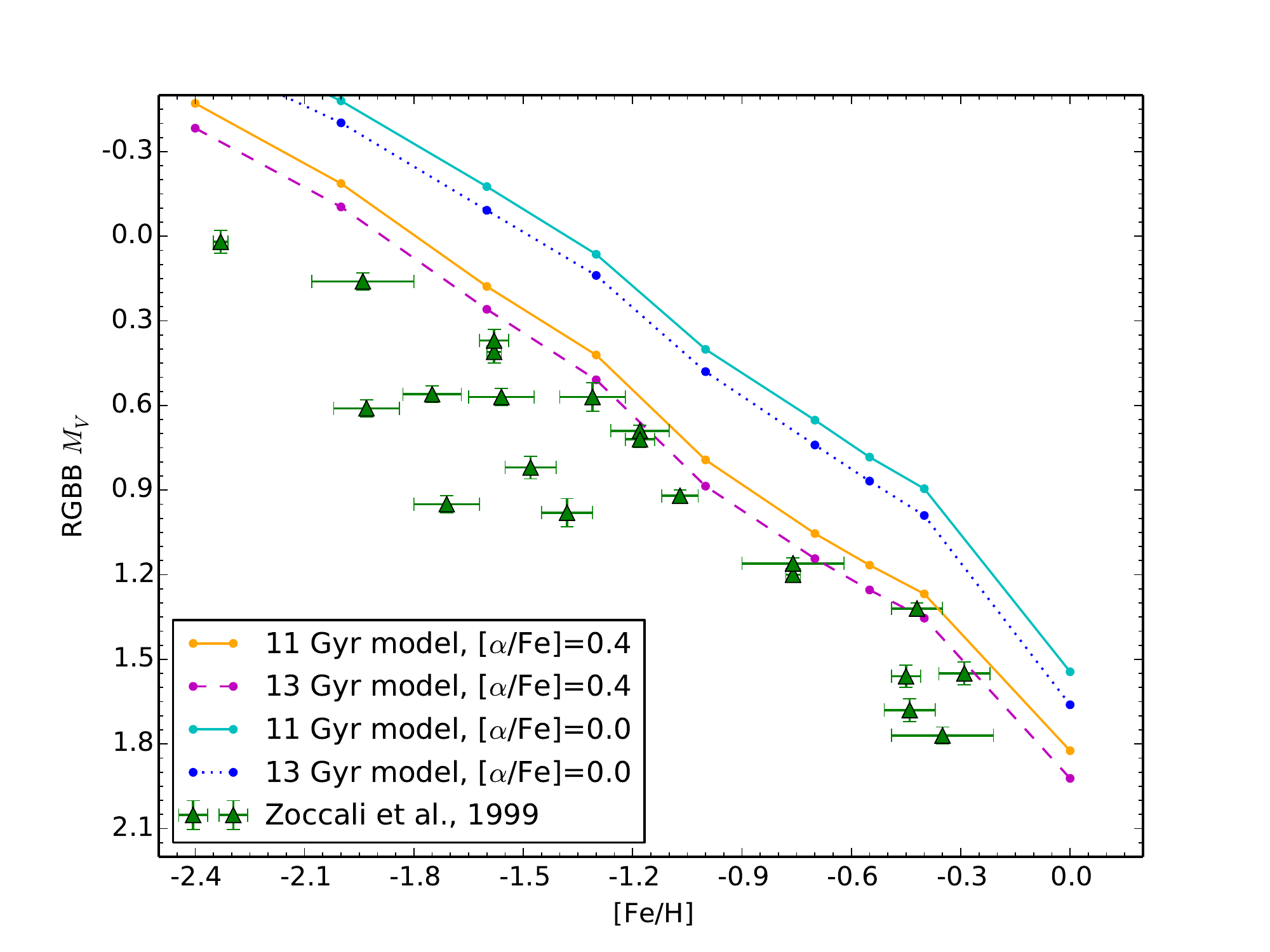}
\caption{Observations of Z1999 are plotted against the DSEP models shown in Figure \ref{NJ}.}
\label{ZJ}
\end{figure}

For completeness, we examine our models' fits to the Z1999 observations as well, and we find that the fit is very poor for a model of any age.  
 The results are shown in Figure \ref{ZJ}. The ``best" fit to Z1999 is, again, our 13 Gyr, $\alpha$-enhanced model, but the spread of the observational data is too large for this result to be meaningful. 

In the previous study comparing DSEP and Z1999, BC2006 determined a theoretical uncertainty of $\sim 0.2$ in their model's predictions of the RGBB magnitude and took this uncertainty into account when stating that the DSEP models were consistent with Z1999's observations. In contrast, our statistics do not include the theoretical uncertainty.

The fact that a 13 Gyr model gives the best fit to N2013's data is reasonable, as globular clusters are thought to form very soon after the Big Bang (13.8 Gyr, \citet{Planck}). That the $\alpha$-enhanced models are a better fit is consistent with our current understanding of globular clusters as well, as observations show that GCs are enhanced in $\alpha$-capture elements (e.g., \citet{Car}). 
Throughout the rest of the paper, we consider only the $\alpha$-enhanced models.

\section{Formal Test of Consistency}
%
%
{ When examining the N2013 GC observations on their own, one can easily recognize that there are clusters whose locations in magnitude-metallicity space are highly unrepresentative of the bulk of the observations. Since it is clear that no single theoretical model could fit the entirety of the observed distribution, we wish to identify the aberrant clusters independently of our models' fits to the data. To do this, we use a Local Outlying Factor (LOF) analysis.  }

The LOF statistical routine is a density-based, model-independent method for identifying the points in a distribution that are furthest from their neighbors \citep{LOF}. The LOF can be done in 4 dimensions, which allows us to take into account both Nataf et al.'s reported magnitude and metallicity errors. A point with an LOF ``$o$-score" close to 1 indicates that it has a low probability of being a density outlier. The $o$-scores for every member of the 72 cluster sample are computed with a standard LOF routine from the package ``DMwR" in the R statistical language \citep{DMwR}. 
The 9 clusters with the highest LOF $o$-scores, or highest probabilities of being density outliers, are presented in in Table \ref{LOFtable}. 

\begin{table} 
\centering 
\caption{Statistical Quantities for N2013's Data: LOF Method}
\begin{tabular}{ l  l c c c c}  
\hline\hline
 { Cluster } & {$o$-score} &{LOF Rank}&[Fe/H]& Sample & $\chi^2$ Rank \\ \hline 
NGC6254	& 1.522			&1		& -1.57	&gold	 & 1 \\
NGC6681	& 1.436			&2		& -1.62	&gold	 & 7 \\
NGC6218	& 1.346			&3		& -1.33	&gold	 & 4 \\
NGC7099 & 1.259			&4		& -2.33	&silver	 & 2 \\
NGC7078	& 1.259			&4		& -2.33	&gold	 & 6 \\
NGC6426	& 1.259			&4		& -2.26	&silver	 & 14 \\
NGC6341	& 1.259			&4		& -2.35	&gold	 & 3 \\
NGC4590	& 1.259			&5		& -2.27	&silver	 & 8 \\
NGC1904	& 1.227			&6		& -1.58	&silver	 & 64 \\ \hline 
\end{tabular}
\tablecomments{The 4D LOF routine identifies the most anomalous clusters based on density. Results are model-independent. The $\chi^2$ rank is a number assigned to indicate where among the $\chi^2$ outliers the named cluster appeared (e.g. $\chi^2$ rank = 2 indicates that this was the second most anomalous cluster according to the $\chi^2$ routine). }
\label{LOFtable}
  \end{table}

\begin{figure} 
\centering
\includegraphics[width=\linewidth]{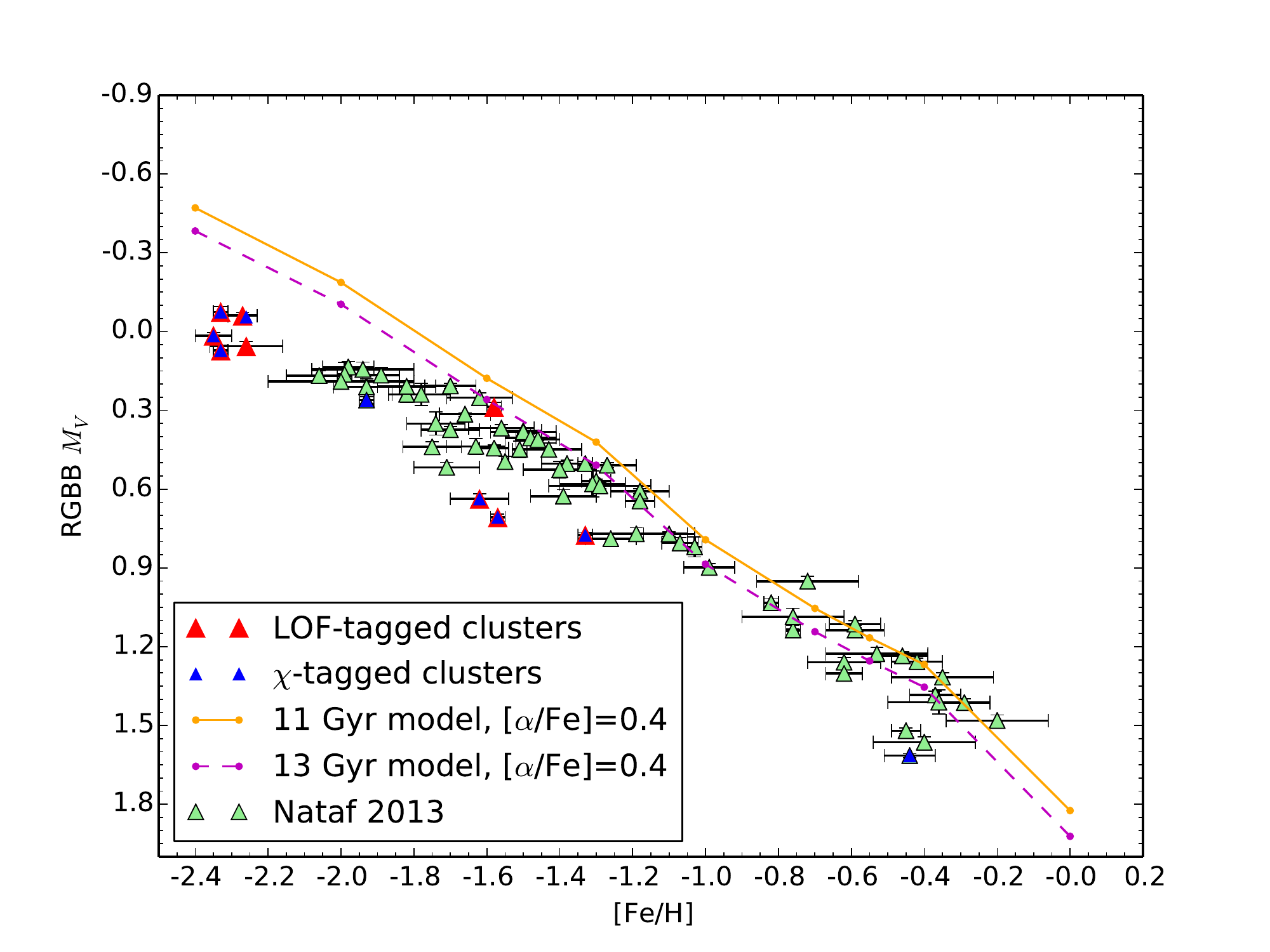}
\caption{The 9 most discrepant clusters determined by the $\chi^2$ test and the 9 most discrepant clusters presented in Table \ref{LOFtable} are shown against the full observational sample and the 11 and 13 Gyr $\alpha$-enhanced models. Clusters indicated with a blue marker on top of a red one are members of both outlier groups. }
\label{selected}
\end{figure}

{
Throughout the rest of our analysis, we refer to cluster observations with high LOF scores as ``anomalous clusters." 
We wish to emphasize that these anomalous clusters are not defined as such because many are highly discrepant with our models---a true finding, and one which will be discussed in more detail later---but fall into this category because they are mathematical outliers determined by a model-independent statistical test. 

An important note should be made regarding NGC7099, NGC7078, NGC6426, NGC6341, and NGC4590, which comprise the clump of clusters in the lowest-metallicity region of the distribution (see  the set of five red clusters shown at [Fe/H]$\approx -2.3$ in Figure \ref{selected} or the upper lefthand corner of Figure \ref{NJ}). 
The tagging of this population by the LOF routine is likely an artifact of the routine's assumption that the entire sample region is well-populated. Since this is not the case,
the comparative isolation of the ultra low-metallicity clusters may artificially inflate their LOF scores. For this reason, we do not include the lowest-metallicity clusters in our references to ``anomalous clusters;" this phrase will be used strictly in reference to NGC6254, NGC6681, NGC6218, and NGC1904. 

Because the routine identifies all of the highly metal-poor clusters as outliers with respect to the sample as whole, it is critical to assess deviations from the observational sample by other means. We proceed using a series of model-dependent $\chi^2$ tests and cross-compare the least representative clusters identified by each method.
}

%
%
The rigorous (dis)agreement of a model with observations is quantified by computing 
\begin{equation}
\chi^2 = \sum^{72}_i \frac{    (M_{V_{i,t}} -  M_{V_{i,o}})^2  (\text{[Fe/H]}_{i,t} - \text{[Fe/H]}_{i,o} )^2   }{  [ (M_{V_{i,t}} -  M_{V_{i,o}})^2 + (\text{[Fe/H]}_{i,t} - \text{[Fe/H]}_{i,o} )^2 ] \sigma_i^2} 
\end{equation}
\begin{equation*}
\text{ with } \sigma_i=\sqrt{\delta_{\text{obs}_i}^2 + \delta_{ \text{[Fe/H]}_i}^2 + \delta_{\text{dist}_i}^2}, 
\end{equation*}
{ where ``t'' subscripts indicate theoretical values, ``o'' subscripts indicate observed values. 
The error factor denoted by $\sigma_i$ comprises three observational uncertainties. The quantity $\delta_{\text{obs}}$ refers to Nataf et al.'s reported magnitude error, and $\delta_{\text{[Fe/H]}}$ refers to their metallicity uncertainty. The quantity $\delta_{\text{dist}}$ is the error imparted by uncertainty in the distance modulus when transforming to absolute magnitudes. For this contribution, we adopt a uniform uncertainty of 0.1 magnitudes. The selection of this value has observational motivation. By analyzing reports of distance moduli in the literature, we find that 0.1 magnitudes is
a reasonable estimate for differences in distance moduli recorded by observers of the same globular cluster. We adopt this uncertainty for all subsequent $\chi^2$ analyses as well.   
}

A reduced $\chi^2$ score of 1.38 is obtained for the fit of a 13 Gyr, $\alpha$-enhanced model to N2013's entire sample.
This corresponds to a $p$-score of 0.0175, indicating that there exists a $\sim$2\% chance of recreating this observational spread with our model. Taking the statistical cutoff for plausible consistency to be $p=0.05$, this score indicates some degree of inconsistency. As was suggested by the LOF routine, however, there are some clusters among the distribution that would not be fit by any model that otherwise fit the bulk of the observations.
 It is thus reasonable to suspect that merely a few cluster observations could be the source of the $\chi^2$ test's failure. Indeed, this is the case. 

{
 For this test, we note the number of observations whose contributions to the total $\chi^2$ score are greater than 5.0. Clusters with higher $\chi^2$ contributions, or $\chi^2_i$ scores, are more statistically discrepant. We refer to clusters with high $\chi^2_i$ contributions as ``$\chi$-tagged clusters." Among Nataf et al.'s full sample, there are four clusters with $\chi^2_i >  5.0$, but
 the removal of only one (NGC 7099) out of 72 observations is required to push our $p$ score above 0.05. The removal of all four $\chi$-tagged clusters (NGCs 7099, 6254, 6341, and 6681) pushes our $p$-score up to 0.5, indicating a high probability of consistency. 
 
Rather than selectively removing the clusters with high $\chi_i$ contributions, it is a more meaningful assessment to remove the clusters identified as anomalous by the LOF routine.
Table \ref{ptable} shows how the reduced $\chi^2$ and $p$ scores change as the anomalous cluster observations are removed from the data set, in order of most to least discrepant.  
The reduced $\chi^2$ and $p$ scores when NGCs 6254, 6681, 6218, and 1904 are removed are 1.17 and 0.16, respectively.
If, instead, the ultra-low metallicity clusters (NGCs 7099, 7078, 6426, 6341, and 4590) alone are removed, the reduced $\chi^2$ score drops to 1.03, with a $p$ score of 0.41. Removing both the anomalous clusters and the ultra-low metallicity clusters tagged by the LOF routine (9 in total), the reduced $\chi^2$ and $p$ scores improve to 0.813 and 0.86, respectively. 
}

Figure \ref{selected} marks the discrepant clusters according to both methods; the samples largely overlap. Table \ref{LOFtable} also indicates the rankings of discrepancy among aberrant clusters by both methods.

Although we have some motivation to question the LOF's tagging of all ultra low-metallicity clusters, it is clear that the $\chi^2$ test also demonstrates statistical issues with these points, and we know that this is not because of any algorithmic quirks. It is worth investigating possible reasons why the low metallicity region is viewed as discrepant by both methods. 
Since the RGBB is harder to detect at lower metallicities, we may expect points in the lowest regime to have greater uncertainty. We have demonstrated why we are skeptical of N2013's reported error in general, and our skepticism broadens when considering the observations that are most difficult to detect. In this regard, we note that the RGBB magnitude for M92 (NGC6341) reported by \citet{Paust} is 0.17 mag brighter than the value reported by \citet{Nataf}, and so would be more consistent with our models.

This aside, the LOF and $\chi^2$ analyses largely agree, especially regarding the clusters which are most problematic for applying a theoretical fit.

\begin{table} 
\centering 
\caption{Statistical Quantities for N2013's Data: $\chi^2$ Method}
\begin{tabular}{ c  l l r c r}  
\hline\hline
 { Sample} & { Reduced $\chi^2$} &      {$p$ score} &   { $\chi^2_i$}    &    { GC}     & {[Fe/H]} \\ \hline
all           & 1.38                    & 0.0175    &  -                    &  none        & - \\  
-1            &  1.26  	               &  0.68  	&   10.01               &  NGC6254     & -1.57 \\ 
-2            & 1.21	               &  0.12	    &    5.13               &  NGC6681     & -1.62 \\ 
-3            & 1.15	               &  0.18	    &    4.75               &  NGC6218     & -1.33		 \\ 
-4            & 1.17	               &  0.15	    &    0.01          		&  NGC1904     & -1.58	\\ \\ \hline 
\end{tabular}
\tablecomments{Members of the LOF-tagged anomalous cluster group are removed from the sample beginning with the most discrepant and working down. The degree of discrepancy is determined by the individual contribution a data point makes to the $\chi^2$ score ($\chi^2_i$). The 13 Gyr, $\alpha$-enhanced model is used in this test. The uncertainty due to distance assumed is $\delta_{\text{dist}} = 0.10$. }
\label{ptable}
  \end{table}

%
%
 We perform the $\chi$-tagging analysis and removal on N2013's gold sample, as shown in Table \ref{Goldptable}. We obtain a reduced $\chi^2$ score of 1.31 when all 48 clusters are included and the 13 Gyr, $\alpha$-enhanced DSEP model is used---an improvement over the fit to the full sample, but one which still produces a $p$ score slightly above the desired minimum. Among this population, there are three highly discrepant clusters: NGC6254, NGC6341, and NGC6681. The progression of the fit statistics with cluster removal shows that discarding the 
 two anomalous clusters pushes the model into the realm of likely consistency. Once again, these three clusters are among those identified by both the LOF algorithm and the $\chi^2$-tagging routine applied to the full sample, providing further evidence that these data should be reconsidered. 

\begin{table} 
\centering 
\caption{Statistical Quantities for N2013's Data: $\chi^2$ Method: Gold Sample}
\begin{tabular}{ c  l l r c r }  
\hline\hline
 { Sample} & { Reduced $\chi^2$} &  {$p$ score} &   { $\chi^2_i$}   &   { GC}     &{[Fe/H]}\\ \hline
all       &  1.31               &   0.0718     &  -                & none           &-       \\  
-1        &  1.13               &   0.254      &  10.014           & NGC6254        & -1.57  \\  
-2         &  1.04               & 0.40          &  5.134           & NGC6681        & -1.62   \\   \hline
\end{tabular}
\tablecomments{Same as Table \ref{ptable}, but using \citet{Nataf}'s gold sample only.}
\label{Goldptable}
\end{table}

%
%
The RGBB magnitude depends sensitively on the composition of the stars, and there may exist systematic differences between the observed globular cluster metallicity and that used in the stellar models. To investigate this possibility, we perform an analysis in which an artificial  shift of [Fe/H]$=\pm0.1$ dex is applied to the observed values. 

{
Unsurprisingly, when the data are shifted by $-0.1$ dex, the fit to any of our models worsens drastically. The $p$ scores in all cases are functionally zero.
When the full data set is shifted by [Fe/H]$=+0.1$, the fit to our 13 Gyr model improves significantly over the fit to the unmodified data. The fit produces a reduced $\chi^2$ of 0.84, a $p$ score of 0.84, and only NGC 7099 and NGC 6254 (the most discrepant in all other cases) have $\chi^2_i$ contributions greater than $5.0$. When the shift is applied, no clusters need be extracted to produce consistency with the whole observational sample. 
When the shift is applied to the gold sample alone, 
the resulting reduced $\chi^2$ and $p$ scores are 0.81 and 0.83, respectively---a slight improvement over the fit to the whole sample. In this case, only  NGC 6254 is tagged as discrepant. 
}

%
%
{
Shifting the metallicity scale uniformly provides an exercise in examining the effects of a systematic uncertainty in the observed GC metallicity scale. However, there is no evidence that suggests an uncertainty would be systematic.
 To take this into consideration, we also test our 13 Gyr, $\alpha$-enhanced model against an alternative metallicity scale provided by \citet{KraftIvans} (KI). Since Kraft \& Ivans (KI) metallicities were not available for all of N2013's GCs, we have performed the statistical tests with only those clusters for which all information is available. This results in a ``full" sample of 40 clusters, and a ``gold" sample of 27 clusters. 

Because the metallicity values have changed and because uncertainties in metallicity are not available with the KI scale, the LOF scores must be recomputed in three dimensions (magnitude, metallicity, and magnitude error) rather than four. Making these adjustments results in slightly different members and orders among the sets of outlying and anomalous clusters. The clusters with the highest LOF scores ($o$-scores above 1.1) according to the KI metallicity scale are, in descending order, NGCs 6254, 6681, 7078, 4590, 5927, 6218, 6093, and 7099. Separating out the clusters with ultra-low metallicities, we are left with NGCs 6254, 6681, 6218, and 6093  composing the KI anomalous cluster group. NGC 5927 is excluded from both groups because the reduction in sample size leaves it isolated at the high-metallicity end of the spectrum; its high $o$-score is likely artificial as well, but it is not a low-metallicity cluster.

The $\chi^2$ results are given in Tables \ref{KI} and \ref{KIgold}, where the magnitudes and uncertainties of the N2013 samples (whole and gold, respectively) are used with the \citet{KraftIvans} metallicities. 
We remove all of the members of the KI anomalous group  
and compute the statistics for the fit to the whole Kraft \& Ivans sample as well. The reduced $\chi^2$ and $p$ scores with the removal of the anomalous clusters only (sample size 36) are 1.12 and 0.28, respectively.
When the KI anomalous and KI metal-poor groups are removed (sample size 33), the reduced $\chi^2$ and $p$ scores are and 0.69 and 0.91, respectively.
The results are consistent with those gathered using the metallicities of N2013, and the most discrepant clusters among both samples remain the same. 
}

%
%
{
\begin{table}
\centering 
\caption{$\chi^2$ Analysis using \citet{KraftIvans} Metallicity Scale}
\begin{tabular}{l c c c r r r}
\hline \hline
{Sample}  & {$\chi^2_R$}  & $p$ score   &{ $\chi_i^2 $}   & GC removed  & N2013   & K\&I  \\ \hline 
 0        & 1.59          & 0.01    &  -              & -           &   -     &  -     \\ 

-1        & 1.37           & 0.06    &  10.01          & NGC6254     & -1.57   & -1.48 \\
-2        & 1.27           & 0.12    &  5.13           & NGC6681     & -1.62   & -1.60 \\ 
-3        &  1.18          & 0.21    & 4.75            & NGC6218     &  -1.33  &  -1.34  \\
-4        &  1.12          & 0.28    & 3.26            & NGC6093     &  -1.75  &  -1.76   \\ \hline
\end{tabular}
\tablecomments{The reduced $\chi^2$ score is computed using the entire observational sample of N2013 but adopting the cluster metallicities reported in \citet{KraftIvans}. Clusters for which \citet{KraftIvans} metallicities were not available are removed from the sample, leaving 40 clusters total. The column marked ``N2013" contains the [Fe/H] values reported by N2013; ``K\&I" contains the values from \citet{KraftIvans}. The age of the model used is 13 Gyr. The distance uncertainty adopted is $\delta_{\text{dist}}=0.10$. The $\alpha$-enhancement is $+0.4$.}
\label{KI}
  \end{table}

\begin{table} 
\centering 
\caption{$\chi^2$ Analysis using \citet{KraftIvans} Metallicity Scale: Gold Sample}
\begin{tabular}{l c l r r r r}
\hline \hline
{Sample}  & {$\chi^2_R$}  & $p$ score   &{ $\chi_i^2 $}   & GC removed  & N2013   & K\&I  \\ \hline 
 0        & 1.34          & 0.11    &  -              & -           &   -     &  -     \\ 
-1        & 1.01          & 0.45    &  10.01          & NGC6254     & -1.57   & -1.48 \\
-2        & 0.84          & 0.69    &  5.13           & NGC6681     & -1.62   & -1.60 \\ 
-3        &  0.68          & 0.88    & 4.75            & NGC6218     &  -1.33  &  -1.34  \\
-4        &  0.56          & 0.95    & 3.26            & NGC6093     &  -1.75  &  -1.76   \\ \hline
\end{tabular}
\tablecomments{Same as Table \ref{KI}, but for N2013's gold sample. \citet{KraftIvans} metallicities are available for 27 gold-sample clusters.}
\label{KIgold}
  \end{table}
} 
%
%
\subsection{Quantifying the Low-Metallicity Discrepancy}
\begin{figure} 
\centering
\includegraphics[width=\linewidth]{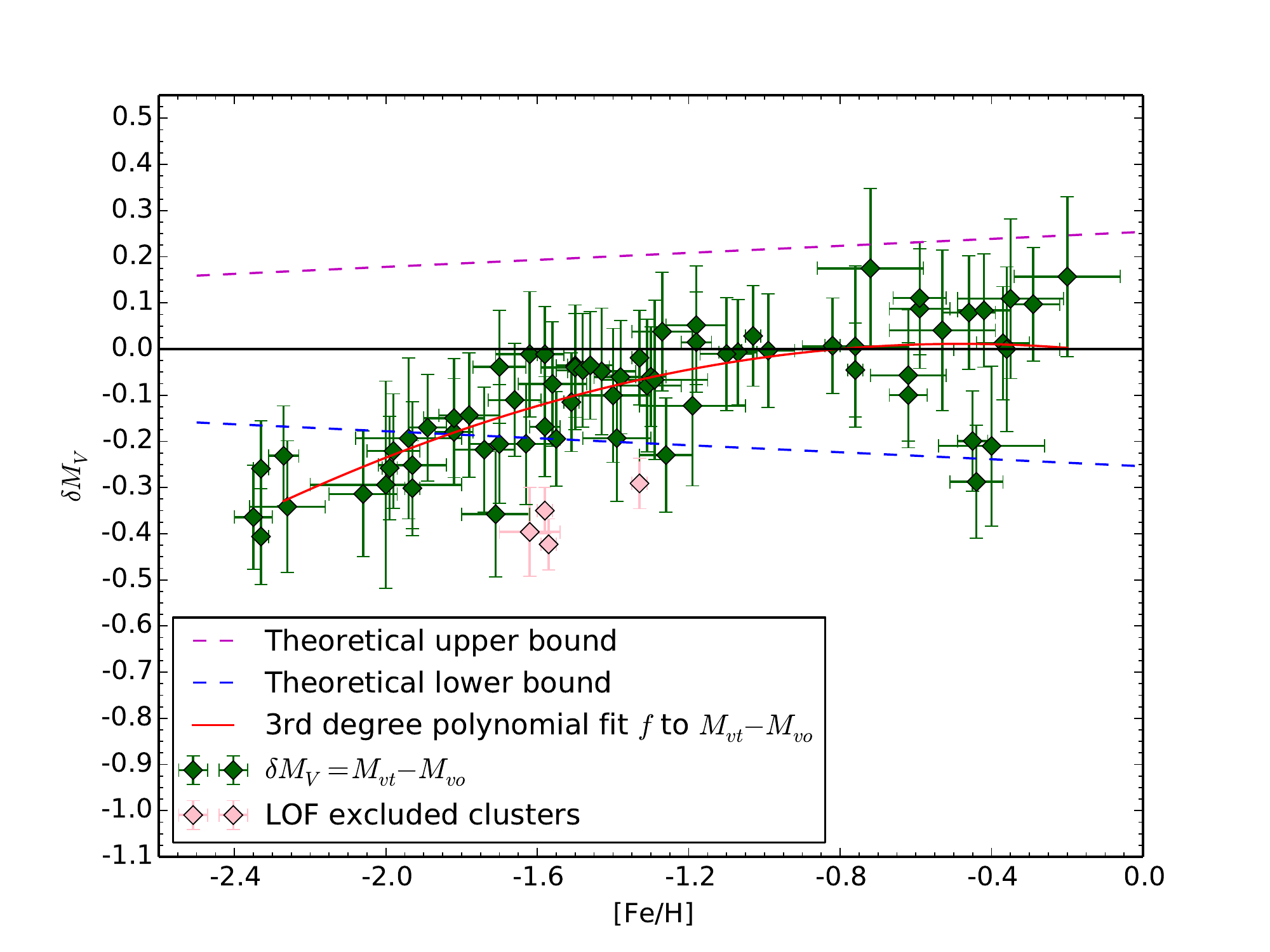}
\caption{The differences between the theoretical and observed magnitudes are shown. Error bars are observational. Theoretical uncertainties that take the computational confidence of \citet{Chab06} into account are superimposed as the upper and lower dashed lines. The trend line (red) is a cubic polynomial fit by a least-squares routine. The fit is computed { without} taking the 7 density outliers into consideration; they are added after and shown as pink squares.}
\label{error}
\end{figure}

Having considered the effects of multiple systematic issues, we now examine the existing trend of increased discrepancy among the lowest metallicity clusters in more detail. 
For each cluster observation, we obtain an associated theoretical value by linearly interpolating between the the RGBB magnitudes from the theoretical points with metallicities closest to the observed cluster's metallicity.
Figure \ref{error} shows the difference in magnitude $\delta M_V = M_{V,\text{model}} - M_{V,\text{observed}}$ as a function of [Fe/H]. Error bars are the same as $\sigma$ in Equation 1.

Figure \ref{error} also includes a theoretical uncertainty, $\delta_{\text{theory}}^2$, which represents the 68\% confidence limit quoted in BC2006. Because the uncertainties reported in BC2006 increase with increasing metallicity, the net theoretical uncertainty widens with increasing metallicity as well. This runs counter to the increase in discrepancy. For metallicities above [Fe/H]= -1.6, nearly all of the observations are contained within the boundaries of the theoretical uncertainties, and the opposite is true for metallicities below this.

{
We quantify the trend by fitting a cubic polynomial to the differences via a maximum likelihood routine. To construct the most accurate mean trend line, 
it is important that we {\it exclude} the four anomalous clusters. 
The clusters excluded from the polynomial fit are shown as pink squares in figure \ref{error} (see also Table \ref{ptable}). The trend line is shown in red. 
}

%
%
In summary, our best-fitting model is shown to agree with N2013's data over the metallicity range [Fe/H]=(0,-1.5) dex. Among the most metal poor clusters, there is the largest disagreement.
{
A few clusters in the more metal-rich regime, however, are anomalous and inconsistent with standard stellar models.}
This may indicate the existence of clusters with, for example, enhanced helium abundance. 

We note that rigid tests of statistical consistency rely heavily on the reported error, and that the most metal-poor observations are the primary sources of discrepancy.
Taking this information into consideration, along with the trend quantified among the residuals, we conclude that, among clusters with metallicities greater than [Fe/H]$=-1.5$, { our model produces a  good fit}. It is unambiguous, however, that DSEP does not fit the most metal-poor selection of GC observations.
However, we reiterate that these are also the clusters for which it is observationally most difficult to measure RGBB magnitudes.

\subsection{A Range of GC Ages}
In the previous analyses, we considered a model of singular age.
In actuality, GCs are not of uniform age, but instead span a small range of ages. As a supplement to performing statistical tests on the best-fitting model with a singular age of 13 Gyr, we examine the statistics of fitting multiple theoretical ages to the N2013 data.

 We consider the sample's fit to a grid of isochrones ranging from 9 to 13 Gyr in increments of 2 Gyr. 
 We proceed using the $\chi^2$ test as described above, including Nataf's magnitude and metallicity errors as well as the 0.1 magnitude error from the distance transformation. Instead of computing the distances of every N2013 point to each curve in the series, however, we consider only the distances of each observation from the age curve to which it is closest.
 Although this necessarily constructs the ideal scenario, we are interested in understanding the best possible fit of our models to observations.
 We exclude ages above $\sim$14 Gyr, as they are not realistic given our current understanding of the age of the universe. This has the effect of fitting every abnormally old cluster to the 13 Gyr isochrone, even though they would be better fit by an unrealistically old model. This necessarily worsens the statistics for that population.

We compute $\chi^2$ scores for each of the subpopulations selected according to age, which are then summed and reduced by the total number of degrees of freedom to obtain the global fit. 
The total reduced $\chi^2$ score for the fit of our multi-age model to N2013's observations is $0.94$, and 3 clusters with $\chi^2_i > 5.0$ are identified: NGC6254, NGC7099, and NGC6341. The $p$ score for the total fit is 0.62. 

The intermediate $p$ scores of each of the age subsets, with the exception of the 13 Gyr group, are all better than those produced by any single-age fit. In fact, they indicate over-fitting ($p$ scores greater than 0.999) in some cases---a product of the contrived nature of this experiment. In short, we find that allowing for multiple models with ages not older than 13 Gyr provides a better fit to the whole sample than does a single 13 Gyr model, but not with great significance, and the discrepant clusters remain the same.

\section{Difference in Magnitude between the RGBB and SGB}

Our comparison to the observed data until now has used the heterogeneous distance moduli reported by N2013. At the expense of requiring an additional theoretically-calculated magnitude, however, we can remove the distance uncertainty by examining the difference in magnitude between the RGBB and sub-giant branch (SGB). Other authors have used the MSTO as a reference magnitude.  However, it is  difficult  to determine the MSTO magnitude in observed data, as the MSTO region is nearly vertical in the color--magnitude diagram. In contrast, the SGB region is nearly horizontal, and so its magnitude is robustly determined in observational data.  The color of the MSTO is well defined, and we define the SGB magnitude to be the magnitude of the point on the SGB which is 0.05 redder than the turn-off.  We elect to use the MSTO colors reported by N2013 (for 48 of the 72 clusters) and then for each of these clusters used a series of binning routines to estimate the SGB magnitudes for each GC.

\begin{figure} 
\centering
\includegraphics[width=\linewidth]{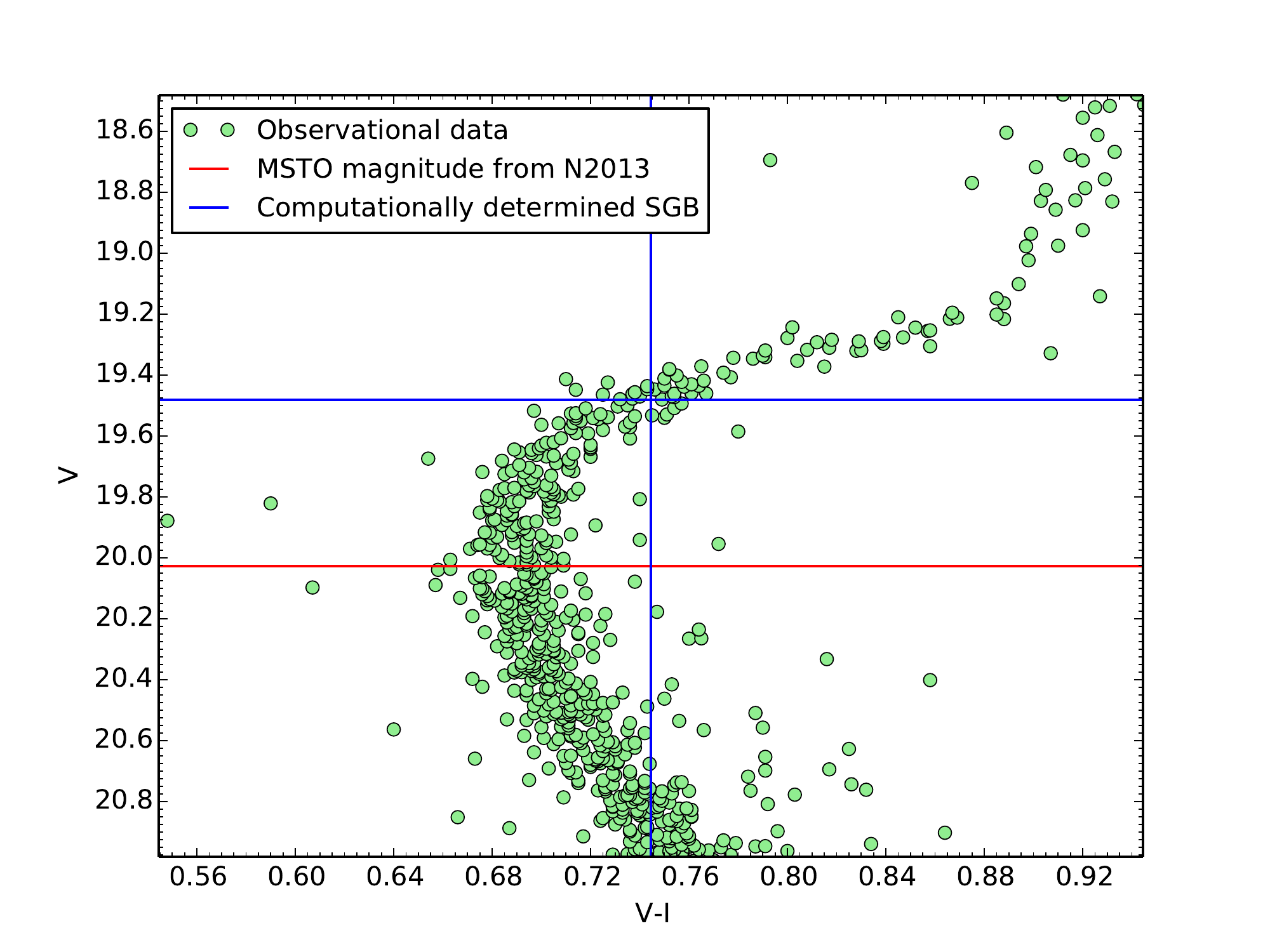}
\caption{An example of the raw cluster data (NGC6584) with N2013's reported MSTO magnitude and the computationally determined SGB magnitudes.}
\label{Vsgb}
\end{figure}

As an example, the raw data from NGC6584 is shown in Figure \ref{Vsgb} along with the points marking the reported MSTO and the algorithm's determination of the SGB magnitude. The V-magnitudes of the SGB are combined with N2013's RGBB V-magnitudes to compute $\Delta V = V_{\text{SGB}} - V_{\text{RGBB}} $.

\begin{figure} 
\centering
\includegraphics[width=\linewidth]{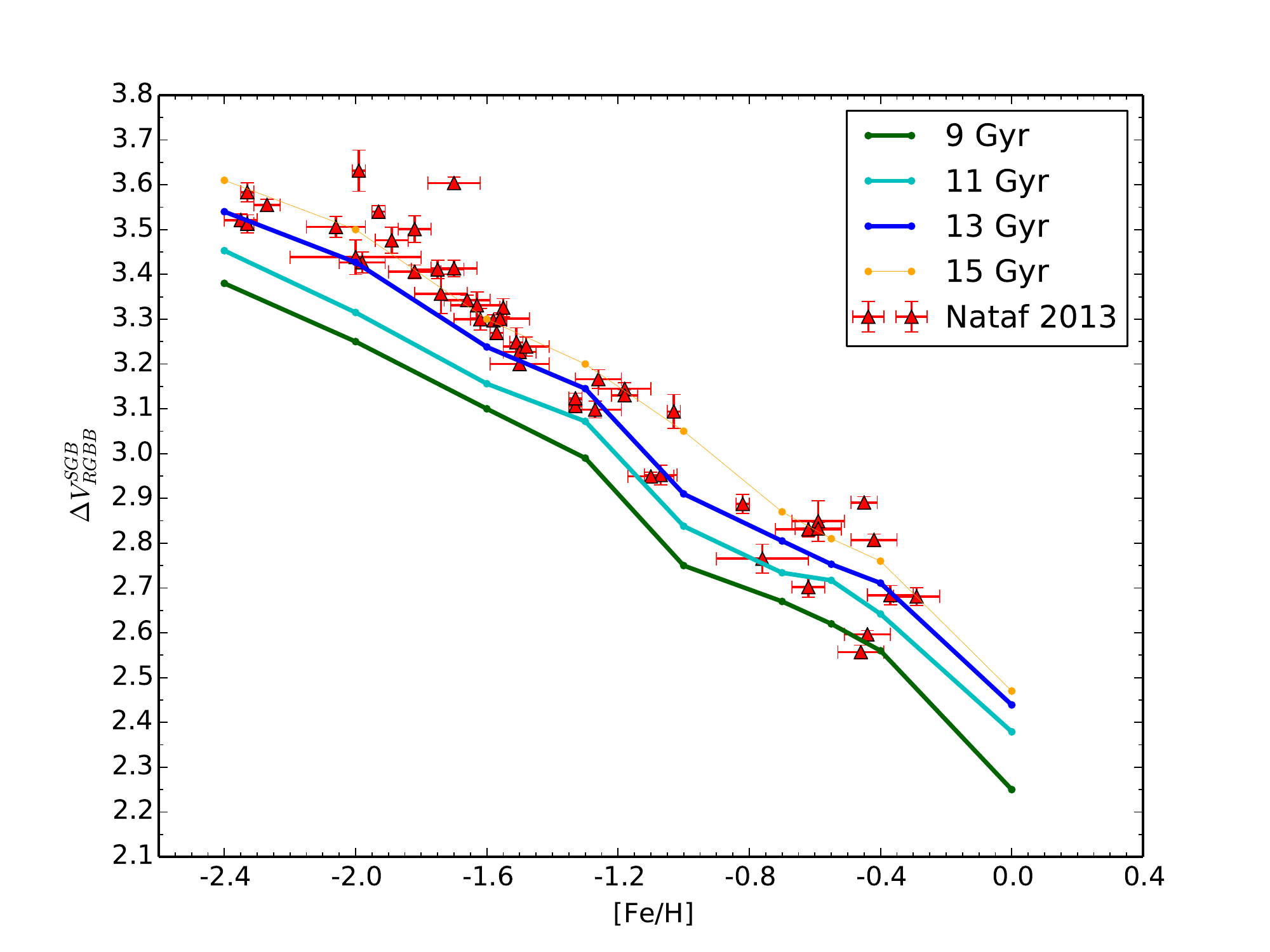}
\caption{The magnitude differences between the SGB and RGBB are shown as a function of metallicity, superimposed on a grid of isochrones with ages as shown.}
\label{agegrid}
\end{figure}

To compare our predicted $\Delta V$ magnitudes for a range of ages to the observations, Figure \ref{agegrid} shows N2013's observations (only those for which MSTO magnitudes are available) superimposed on a grid of DSEP isochrones ranging from 9 to 15 Gyr. 
{
Observed values of $\Delta V$ are systematically larger than predicted values, and the most discrepant clusters have changed. The two most obvious outliers according to the new metric are NGC 6397 and NGC 6656. As was the case in previous analyses, problems with the theoretical models are most apparent at lower metallicities. 
}

As in the previous section, for each observational point, the minimum distance of the point to the nearest isochrone is computed. Higher resolution is synthesized by interpolating between the 2 Gyr tracks to obtain increments of 0.5 Gyr. 
By this method, we assign an age, within 0.5 Gyr, to each of these 48 clusters. Ages are given in Table \ref{agetable} for clusters with fit ages $\le 13.0$ Gyr.

We reiterate that clusters with assigned ages ${\ge}$14 Gyr arouse suspicion, yet the majority of this cluster population appears to be best fit by ages older than 13.5 Gyr. To examine a multi-age fit that is physically reasonable, we reassign all ages greater than 13.0 Gyr to the 13 Gyr isochrone and perform a 
$\chi^2$ test. The test returns a reduced $\chi^2$ score of 2.99 when no ages greater than 13 Gyr are permitted. This is a significantly worse score than the one obtained in section 5.1, but this is to be expected given the $\chi^2$ test's reliance on the observer's error bars and the drastic reduction in uncertainty imparted by removing the distance error. If, for example, the error bars were doubled, the reduced $\chi^2$ score would drop to 1.50 with an accompanying $p=0.015$. 

{
 Reduced  $\chi^2$ and $p$ scores are computed for each subpopulation's fit to age curves spaced 0.5 Gyr apart, and then the non-reduced $\chi^2$ scores are summed and divided by the total degrees of freedom to obtain the final reduced $\chi^2$ and $p$ scores.
As one could imagine, highly discrepant clusters are tagged in the 13 Gyr group (only). Among this population, 7 clusters with $\chi^2_i > 5.0$ are identified: NGCs 6656, 6397, 6388, 6144, 6809, 5268, and 6752. It should be noted that these clusters do not overlap significantly with clusters tagged as aberrant by other tests. There are a few reasons for this.
 First, the age curves in this test are constructed from measurements of the SGB brightness, which is subject to uncertainty in the estimate of the SGB magnitude and, by extension, uncertainty in the estimate of the color of the MSTO.
  These uncertainties are distinct from the uncertainties in the RGBB magnitude, which have remained constant throughout all previous tests. 
 The aberrant clusters, from the perspective of this method, will skew towards those whose theoretical and predicted SGB magnitudes disagree---a feature we have not assessed until this point.
 We also note that the N2013 subpopulation for which MSTO magnitudes and metallicities are available only contains 48 of the original 72 clusters.
}

\begin{table}  
\centering 
\caption{Clusters with Fitted Ages $\le 13$ Gyr}
\begin{tabular}{c c c}
\hline\hline
 {Cluster} 	& {Age (Gyr) } 	& {[Fe/H]} \\ \hline
 NGC1261	&  12.0		& -1.27	\\	
 NGC5904   	&  12.0   	& -1.33	\\ 
 NGC6218   	&  12.0  	& -1.33	 \\ 
 NGC6304   	&  13.0   	& -0.37	\\
 NGC6341   	&  13.0 	& -2.35	\\ 
 NGC6352  	&  10.5    	& -0.62	\\ 
 NGC6362   	&  12.0    	& -1.07	\\ 
 NGC6441   	&  9.5    	& -0.44	\\
 NGC6496   	&  9.0     	& -0.46	\\ 
 NGC6584   	&  12.5   	& -1.5	\\ 
 NGC6652   	&  11.0   	& -0.76	 \\ 
 NGC6723   	&  12.0   	& -1.1	\\ 
 NGC7099   	&  12.5   	& -2.33	\\ \hline
\end{tabular}
\tablecomments{Ages are assigned based on the $\Delta V^{\text{SGB}}_{\text{RGBB}}$ isochrones. Clusters are members of N2013's 48-cluster subsample. }
\label{agetable}  \end{table}

~~\\
\section{Summary and Future Work}
Roughly a decade ago, BC2006 found that the DSEP stellar models showed good agreement with the observed magnitudes of the RGBB in a sample of 28 Galactic globular clusters presented by Z1999.  Since then, studies by \citet{Di} and \citet{Cas} found that stellar models calculated by other groups were inconsistent with more recent observational data. We have presented a new set of synthetic RGBB magnitudes calculated with an improved version of DSEP.
We find that our latest RGBB magnitudes agree, within the error imparted by microphysical differences, with those predicted by the YY models \citep{YY}, and VR models \citep{VDB2006}, and to a lesser extent, the BaSTI models \citet{BaSTI} and MESA models \citet{MESA}.

A DSEP model of age 13 Gyr with enhanced $\alpha$-abundance is found to demonstrate the best fit to every data sample we consider.
When we compared our best-fitting model to the extensive observational data set of N2013, we found some disagreement between our models and the observations by several measures. A series of $\chi^2$ tests, an LOF analysis, and a fit to the mean differences between projected and observed magnitudes all reveal that $\sim$5-10\% of the sample is largely discrepant with our model. The same clusters are repeatedly identified as outliers by different routines, and their obvious common feature is low metallicity. 

 We have provided statistical evidence for the identification of outliers, and suggest that unusual cluster properties may be the culprit, especially for those clusters which occur in many of the outlier groups. 

{
We have found that for [Fe/H] $> -1.5$, the observed RGBB magnitudes are well within our estimated theoretical uncertainties, 
indicating that, in order to fit the observations, models would require relatively small corrections that are within the known uncertainties of those models.  In contrast, if the lowest metallicity points ([Fe/H] $\sim$-2.3) are correct, the stellar models will require substantial revisions. It is thus fair to say that the DSEP models provide a { reasonable} fit to observations {\it above the breaking point ([Fe/H] $\ge -1.5$)}. 

In general, when comparing our predicted RGBB absolute magnitudes to the observations, we find that our magnitudes are too bright in the region [Fe/H] $< -1.2$.  However, when looking at the difference in magnitude between the subgiant branch and the RGBB, we find that our theoretical values are typically smaller than the observations. This would suggest that either our RGBB magnitudes are too faint, our subgiant magnitudes are too bright, or that the observed location of the subgiant branch has been subject to some systematic measurement error. 
This issue requires further investigation. 
The essential result is that the two approaches to comparing predicted RGBB magnitudes to observations agree that the discrepancy between the models and the data is largest at the lower metallicities.
}

We can elaborate upon this study in a number of ways in the future. First, it would be instructive to examine the discrepant clusters in both the high and low metallicity regimes---in particular NGC6254, NGC7099, and NGC6681 
---in more detail, both from the observational and theoretical perspectives.  Observationally, independent estimates of the RGBB magnitudes and their errors for the LOF-tagged anomalous population would be informative. From a theoretical standpoint, 
we could learn more by running stellar models tailored to the observed cluster parameters and investigating whether custom stellar models provide better agreement with the data. In addition, we hope to perform the analyses presented in this paper for the RGBB in terms of star number count and compare to the recent work of \citet{N2014}.

Finally, there is a wealth of evidence that many globular clusters contain multiple stellar populations. We hope to adapt our techniques to facilitate 
 examining the impact of this property on the predicted location of the RGBB.

~~\\
This work is supported by grant AST-1211384 from the National Science Foundation. We thank the statistical reviewer and referee for their instructive comments. 
\bibliographystyle{apj}
\bibliography{RGBB_Sept2015}

\begin{thebibliography}{}
\expandafter\ifx\csname natexlab\endcsname\relax\def\natexlab#1{#1}\fi

\bibitem[{{Adelberger} {et~al.}(1998){Adelberger}, {Austin}, {Bahcall},
  {Balantekin}, {Bogaert}, {Brown}, {Buchmann}, {Cecil}, {Champagne}, {de
  Braeckeleer}, {Duba}, {Elliott}, {Freedman}, {Gai}, {Goldring}, {Gould},
  {Gruzinov}, {Haxton}, {Heeger}, {Henley}, {Johnson}, {Kamionkowski},
  {Kavanagh}, {Koonin}, {Kubodera}, {Langanke}, {Motobayashi}, {Pandharipande},
  {Parker}, {Robertson}, {Rolfs}, {Sawyer}, {Shaviv}, {Shoppa}, {Snover},
  {Swanson}, {Tribble}, {Turck-Chi{\`e}ze}, \& {Wilkerson}}]{Adel}
{Adelberger}, E.~G., {Austin}, S.~M., {Bahcall}, J.~N., {et~al.} 1998, Reviews
  of Modern Physics, 70, 1265

\bibitem[{{Bergbusch} \& {VandenBerg}(2001)}]{Bergbusch}
{Bergbusch}, P.~A., \& {VandenBerg}, D.~A. 2001, \apj, 556, 322

\bibitem[{{Bjork} \& {Chaboyer}(2006)}]{Chab06}
{Bjork}, S.~R., \& {Chaboyer}, B. 2006, \apj, 641, 1102

\bibitem[{{Bressan} {et~al.}(2012){Bressan}, {Marigo}, {Girardi}, {Salasnich},
  {Dal Cero}, {Rubele}, \& {Nanni}}]{CMD}
{Bressan}, A., {Marigo}, P., {Girardi}, L., {et~al.} 2012, \mnras, 427, 127

\bibitem[{Breunig {et~al.}(2000)Breunig, Kriegel, Ng, \& Sander}]{LOF}
Breunig, M.~M., Kriegel, H.-P., Ng, R.~T., \& Sander, J. 2000, SIGMOD Rec., 29,
  93

\bibitem[{{Carretta} {et~al.}(2010){Carretta}, {Gratton}, {Lucatello},
  {Bragaglia}, {Catanzaro}, {Leone}, {Momany}, {D'Orazi}, {Cassisi},
  {D'Antona}, \& {Ortolani}}]{Car}
{Carretta}, E., {Gratton}, R.~G., {Lucatello}, S., {et~al.} 2010, \apjl, 722,
  L1

\bibitem[{{Cassisi} {et~al.}(1997){Cassisi}, {degl'Innocenti}, \&
  {Salaris}}]{CDS}
{Cassisi}, S., {degl'Innocenti}, S., \& {Salaris}, M. 1997, \mnras, 290, 515

\bibitem[{{Cassisi} {et~al.}(2011){Cassisi}, {Mar{\'{\i}}n-Franch}, {Salaris},
  {Aparicio}, {Monelli}, \& {Pietrinferni}}]{Cas}
{Cassisi}, S., {Mar{\'{\i}}n-Franch}, A., {Salaris}, M., {et~al.} 2011, \aap,
  527, A59

\bibitem[{{Cassisi} \& {Salaris}(1997)}]{CS97}
{Cassisi}, S., \& {Salaris}, M. 1997, \mnras, 285, 593

\bibitem[{{Cordier} {et~al.}(2007){Cordier}, {Pietrinferni}, {Cassisi}, \&
  {Salaris}}]{BaSTI}
{Cordier}, D., {Pietrinferni}, A., {Cassisi}, S., \& {Salaris}, M. 2007, \aj,
  133, 468

\bibitem[{{Demarque} {et~al.}(2004){Demarque}, {Woo}, {Kim}, \& {Yi}}]{YY}
{Demarque}, P., {Woo}, J.-H., {Kim}, Y.-C., \& {Yi}, S.~K. 2004, \apjs, 155,
  667

\bibitem[{{Di Cecco} {et~al.}(2010){Di Cecco}, {Bono}, {Stetson},
  {Pietrinferni}, {Becucci}, {Cassisi}, {Degl'Innocenti}, {Iannicola}, {Prada
  Moroni}, {Buonanno}, {Calamida}, {Caputo}, {Castellani}, {Corsi}, {Ferraro},
  {Dall'Ora}, {Monelli}, {Nonino}, {Piersimoni}, {Pulone}, {Romaniello},
  {Salaris}, {Walker}, \& {Zoccali}}]{Di}
{Di Cecco}, A., {Bono}, G., {Stetson}, P.~B., {et~al.} 2010, \apj, 712, 527

\bibitem[{{Dotter} {et~al.}(2008){Dotter}, {Chaboyer}, {Jevremovi{\'c}},
  {Kostov}, {Baron}, \& {Ferguson}}]{Dotter}
{Dotter}, A., {Chaboyer}, B., {Jevremovi{\'c}}, D., {et~al.} 2008, \apjs, 178,
  89

\bibitem[{{Ferraro} {et~al.}(1999){Ferraro}, {Messineo}, {Fusi Pecci}, {de
  Palo}, {Straniero}, {Chieffi}, \& {Limongi}}]{Ferraro}
{Ferraro}, F.~R., {Messineo}, M., {Fusi Pecci}, F., {et~al.} 1999, \aj, 118,
  1738

\bibitem[{{Fusi Pecci} {et~al.}(1990){Fusi Pecci}, {Ferraro}, {Crocker},
  {Rood}, \& {Buonanno}}]{Pecci}
{Fusi Pecci}, F., {Ferraro}, F.~R., {Crocker}, D.~A., {Rood}, R.~T., \&
  {Buonanno}, R. 1990, \aap, 238, 95

\bibitem[{{Hauschildt} {et~al.}(1999){Hauschildt}, {Allard}, \&
  {Baron}}]{Phoenix}
{Hauschildt}, P.~H., {Allard}, F., \& {Baron}, E. 1999, \apj, 512, 377

\bibitem[{{Irwin}(2012)}]{FreeEOS}
{Irwin}, A.~W. 2012, {FreeEOS: Equation of State for stellar interiors
  calculations}, Astrophysics Source Code Library, ascl:1211.002

\bibitem[{{Kraft} \& {Ivans}(2003)}]{KraftIvans}
{Kraft}, R.~P., \& {Ivans}, I.~I. 2003, \pasp, 115, 143

\bibitem[{Marta {et~al.}(2008)Marta, Formicola, Gy\"urky, Bemmerer, Broggini,
  Caciolli, Corvisiero, Costantini, Elekes, F\"ul\"op, Gervino, Guglielmetti,
  Gustavino, Imbriani, Junker, Kunz, Lemut, Limata, Mazzocchi, Menegazzo,
  Prati, Roca, Rolfs, Romano, Alvarez, Somorjai, Straniero, Strieder, Terrasi,
  Trautvetter, \& Vomiero}]{Marta}
Marta, M., Formicola, A., Gy\"urky, G., {et~al.} 2008, Phys. Rev. C, 78, 022802

\bibitem[{{Michaud} {et~al.}(2010){Michaud}, {Richer}, \& {Richard}}]{MRR}
{Michaud}, G., {Richer}, J., \& {Richard}, O. 2010, \aap, 510, A104

\bibitem[{{Nataf}(2014)}]{N2014}
{Nataf}, D.~M. 2014, \mnras, 445, 3839

\bibitem[{{Nataf} {et~al.}(2013){Nataf}, {Gould}, {Pinsonneault}, \&
  {Udalski}}]{Nataf}
{Nataf}, D.~M., {Gould}, A.~P., {Pinsonneault}, M.~H., \& {Udalski}, A. 2013,
  \apj, 766, 77

\bibitem[{{Paust} {et~al.}(2007){Paust}, {Chaboyer}, \& {Sarajedini}}]{Paust}
{Paust}, N.~E.~Q., {Chaboyer}, B., \& {Sarajedini}, A. 2007, \aj, 133, 2787

\bibitem[{{Paxton} {et~al.}(2013){Paxton}, {Cantiello}, {Arras}, {Bildsten},
  {Brown}, {Dotter}, {Mankovich}, {Montgomery}, {Stello}, {Timmes}, \&
  {Townsend}}]{MESA}
{Paxton}, B., {Cantiello}, M., {Arras}, P., {et~al.} 2013, \apjs, 208, 4

\bibitem[{{Pietrinferni} {et~al.}(2010){Pietrinferni}, {Cassisi}, \&
  {Salaris}}]{Piet10}
{Pietrinferni}, A., {Cassisi}, S., \& {Salaris}, M. 2010, \aap, 522, A76

\bibitem[{{Pietrinferni} {et~al.}(2006){Pietrinferni}, {Cassisi}, {Salaris}, \&
  {Castelli}}]{Piet06}
{Pietrinferni}, A., {Cassisi}, S., {Salaris}, M., \& {Castelli}, F. 2006, \apj,
  642, 797

\bibitem[{{Planck Collaboration}(2013)}]{Planck}
{Planck Collaboration}. 2013, VizieR Online Data Catalog, 8091, 0

\bibitem[{{Pollard} {et~al.}(1994){Pollard}, {Cottrell}, \& {Lawson}}]{Pol}
{Pollard}, K.~R., {Cottrell}, P.~L., \& {Lawson}, W.~A. 1994, \mnras, 268, 544

\bibitem[{{Riello} {et~al.}(2003){Riello}, {Cassisi}, {Piotto}, {Recio-Blanco},
  {De Angeli}, {Salaris}, {Pietrinferni}, {Bono}, \& {Zoccali}}]{Riello}
{Riello}, M., {Cassisi}, S., {Piotto}, G., {et~al.} 2003, \aap, 410, 553

\bibitem[{{Sarajedini} {et~al.}(2007){Sarajedini}, {Bedin}, {Chaboyer},
  {Dotter}, {Siegel}, {Anderson}, {Aparicio}, {King}, {Majewski},
  {Mar{\'{\i}}n-Franch}, {Piotto}, {Reid}, \& {Rosenberg}}]{Sara}
{Sarajedini}, A., {Bedin}, L.~R., {Chaboyer}, B., {et~al.} 2007, \aj, 133, 1658

\bibitem[{Torgo(2010)}]{DMwR}
Torgo, L. 2010, Data Mining with R, learning with case studies (Chapman and
  Hall/CRC)

\bibitem[{{Troisi} {et~al.}(2011){Troisi}, {Bono}, {Stetson}, {Pietrinferni},
  {Weiss}, {Fabrizio}, {Ferraro}, {Cecco}, {Iannicola}, {Buonanno}, {Calamida},
  {Caputo}, {Corsi}, {Dall'Ora}, {Kunder}, {Monelli}, {Nonino}, {Piersimoni},
  {Pulone}, {Romaniello}, {Walker}, \& {Zoccali}}]{Troisi}
{Troisi}, F., {Bono}, G., {Stetson}, P.~B., {et~al.} 2011, \pasp, 123, 879

\bibitem[{{VandenBerg} {et~al.}(2006){VandenBerg}, {Bergbusch}, \&
  {Dowler}}]{VDB2006}
{VandenBerg}, D.~A., {Bergbusch}, P.~A., \& {Dowler}, P.~D. 2006, \apjs, 162,
  375

\bibitem[{{VandenBerg} \& {Clem}(2003)}]{VC}
{VandenBerg}, D.~A., \& {Clem}, J.~L. 2003, \aj, 126, 778

\bibitem[{{Zoccali} {et~al.}(1999){Zoccali}, {Cassisi}, {Piotto}, {Bono}, \&
  {Salaris}}]{Zoc}
{Zoccali}, M., {Cassisi}, S., {Piotto}, G., {Bono}, G., \& {Salaris}, M. 1999,
  \apjl, 518, L49

\end{thebibliography}

\end{document}